\documentclass[preprint2]{emulateapj}

\newcommand{\eps}[1]{\log A_{\rm #1}}
\newcommand{\Eps}{\log\varepsilon}
\newcommand{\kms}{km\,s$^{-1}$}

\newcommand{\kH}{$S_{\rm H}$}
\newcommand{\kh}{$S_{\rm H}$}
\newcommand{\teff}{$T_{\rm eff}$}
\newcommand{\Teff}{T_{\rm eff}}
\newcommand{\eexc}{E_{\rm exc}}
\newcommand{\vt}{\xi_t}

\slugcomment{}

\begin{document}


\shorttitle{NLTE study of F and G dwarfs. I. Stellar atmosphere parameters.} \shortauthors{Sitnova et al.}

\title{\textbf{Systematic NLTE study of the $-2.6 \le$ [Fe/H] $\le 0.2$ F and G dwarfs \\
 in the solar neighbourhood. \\
I. Stellar atmosphere parameters \thanks{Based on
    observations collected at the UCO/Lick observatory, USA and Canada-France-Hawaii Telescope}}}

\author{T. Sitnova}

\affil{Key Laboratory of Optical Astronomy, National Astronomical Observatories, Chinese Academy of Sciences, Beijing 100012, China} \email{sitnova@inasan.ru}
\affil{Institute of Astronomy, Russian Academy of Sciences, RU-119017 Moscow, Russia}
\affil{Lomonosov Moscow State University, RU-119991 Moscow, Russia}

\author{G. Zhao }
\affil{Key Laboratory of Optical Astronomy, National Astronomical Observatories, Chinese Academy of Sciences, Beijing 100012, China}

\author{L. Mashonkina }
\affil{Institute of Astronomy, Russian Academy of Sciences, RU-119017 Moscow, Russia}

\author{Y. Chen}
\affil{Key Laboratory of Optical Astronomy, National Astronomical Observatories, Chinese Academy of Sciences, Beijing 100012, China}

\author{F. Liu}
\affil{Key Laboratory of Optical Astronomy, National Astronomical Observatories, Chinese Academy of Sciences, Beijing 100012, China}
\affil{ Research School of Astronomy and Astrophysics, Australian National University, Canberra, ACT 2611, Australia}

\author{Yu. Pakhomov}
\affil{Institute of Astronomy, Russian Academy of Sciences, RU-119017 Moscow, Russia}

\author{K. Tan}
\affil{Key Laboratory of Optical Astronomy, National Astronomical Observatories, Chinese Academy of Sciences, Beijing 100012, China}

\author{M. Bolte}
\affil{UCO/Lick Observatory, University of California, 1156 High St., Santa Cruz, CA 95064, USA}

\author{S. Alexeeva}
\affil{Institute of Astronomy, Russian Academy of Sciences, RU-119017 Moscow, Russia}

\author{F. Grupp}
\affil{Universitats Sternwarte Munchen, Scheinerstr. 1, D-81679 Munchen, Germany
}
\affil{Max-Planck-Institut fur extraterrestrische Physik, Giessenbachstrasse, D-85748 Garching, Germany
}

\author{J.-R. Shi}
\affil{Key Laboratory of Optical Astronomy, National Astronomical Observatories, Chinese Academy of Sciences, Beijing 100012, China}

\author{H.-W. Zhang}
\affil{Department of Astronomy, School of Physics, Peking University, Beijing 100871, PR China}


\begin{abstract}

We present atmospheric parameters for 51 nearby F and G dwarf and subgiant stars uniformly distributed over the $-2.60 < {\rm [Fe/H]} < +0.20$ metallicity range that is suitable for the Galactic chemical evolution research. Lines of iron in the two ionization stages, Fe~I and Fe~II, were used to derive a homogeneous set of effective temperatures, surface gravities, iron abundances, and microturbulence velocities. Our spectroscopic analyses took advantage of employing high-resolution ($R \ge$ 60\,000) Shane/Hamilton and CFHT/ESPaDOnS observed spectra and non-local thermodynamic equilibrium (NLTE) line formation for Fe~I and Fe~II in the classical 1D model atmospheres. The spectroscopic method was tested in advance with the 20 benchmark stars, for which there are multiple measurements of the infrared flux method (IRFM) effective temperature and their Hipparcos parallax error is less than 10\,\%. We found NLTE abundances from lines of Fe~I and Fe~II to be consistent within 0.06~dex for every benchmark star, when applying a scaling factor of \kH\ = 0.5 to the Drawinian rates of inelastic Fe+H collisions. The obtained atmospheric parameters were checked for each program star by comparing its position in the log~g -- \teff\ plane with the theoretical evolutionary track of given metallicity and $\alpha$-enhancement in the \citet{Yi2004} grid. Our final effective temperatures lie exactly in between the $T_{\rm IRFM}$ scales of \citet{Alonso1996irfm} and \citet{Casagrande2011}, with a mean difference of +46~K and $-51$~K, respectively. NLTE leads to higher surface gravity compared with that for LTE. The shift in log~g is smaller than 0.1~dex for stars with either [Fe/H] $\ge -0.75$, or \teff\ $\le$ 5750~K, or log~g $\ge$ 4.20. 
 NLTE analysis is crucial for the VMP turn-off and subgiant stars, for which the shift in log~g between NLTE and LTE can be up to 0.5~dex.
The obtained accurate atmospheric parameters will be used in the forthcoming papers to determine NLTE abundances of important astrophysical elements from lithium to europium and to improve observational constraints on the chemo-dynamical models of the Galaxy evolution.

\end{abstract}

\keywords{Stars: abundances  -- Stars: atmospheres -- Stars: fundamental parameters} 

\section{Introduction}

In recent decades the construction of large telescopes, the developement of efficient spectrometers, and progress in stellar atmosphere and synthetic spectrum numerical modelling provided a considerable improvement in the quality and quantity of stellar abundance determinations. 
Elemental abundances for the FGK-type stars provide important clues to understand the main processes at play in formation and evolution of the Milky Way (MW). Studies of stellar samples with metallicity [Fe/H]\footnote{In the classical notation, where [X/H] = $\log(N_{\rm X}/N_{\rm H})_{star} - \log(N_{\rm X}/N_{\rm H})_{Sun}$.}   $\ge -1$  showed how trends in various chemical elements can be used to learn the chemodynamical evolution of the Galactic disk \citep{Edvardsson1993A&A...275..101E,2000A&AS..141..491C} and to resolve the thick disk and thin disk 
\citep[][and references therein]{Gratton1996ASPC...92..307G,Fuhrmann1998,Fuhrmann2004,mash_eu,2003MNRAS.340..304R,Mishenina2004,Zhang2006,Adibekyan2013,Bensby2014}.
Studies of very metal-poor (VMP, [Fe/H] $\le -2$) stars unraveled the major processes of chemical enrichment of the Milky Way \citep{1995AJ....109.2757M,Cayrel2004,Zhang2005,Bonifacio2009A&A...501..519B}.

Elemental abundances of stars with different metallicities are the main constraint for Galactic chemodynamical models ( e.g. \citet{2001ApJ...554.1044C,2010A&A...522A..32R,2011MNRAS.414.3231K}). For the Galactic chemical evolution research it would be useful to deal with a homogeneous set of accurate stellar abundances in wide metallicity range, from supersolar down to extremely low iron abundances. 
Large high-resolution spectroscopic surveys were proposed to increase the statistics of observations and to improve a homogeneity of derived stellar chemical abundances over wide metallicity range.
The Apache Point Observatory Galactic Evolution Experiment (APOGEE) \citep{APOGEE2013AJ....146...81Z}, with its $~10^5$ red giants observed in the near-infrared H-band with a spectral resolving power of $R \simeq$ 22\,500, enables users to address numerous Galactic structure and stellar populations issues. 
The Gaia-ESO Survey consortium \citep{2012Msngr.147...25G,2013Msngr.154...47R,Smiljanic2014arXiv1409.0568S} is  obtaining high-quality spectroscopic data for about 10$^5$ stars using FLAMES at the VLT. 
Spectra for a million stars will be acquired by the coming Galactic Archaeology with HERMES (GALAH) survey \citep{2015arXiv150204767D}.

We initiate a new project of deriving a homogeneous set of stellar atmosphere parameters and chemical abundances for
the Galactic field FGK-type stars in the metallicity range $-3 \le$ [Fe/H] $\le +0.3$ that is suitable for a systematic research of Galactic chemical evolution. 
By employing high-resolution spectral observations, deriving accurate stellar atmosphere parameters, and treating the non-local thermodynamic equilibrium (NLTE) line formation for the key chemical species, we attempt to push the accuracy of the abundance analysis to the points where the trends with metallicity could be meaningfully discussed. The selected stars are uniformly distributed in metallicity and they can serve as a calibration sample for the existing and coming large-scale stellar surveys, such as the LAMOST Experiment for Galactic Understanding and Exploration \citep[LEGUE][]{2012RAA....12..735D}, designed to determine from low-resolution spectra elemental abundances for hundreds of thousands to millions of stars, spread over much larger distances than ever before. 

 The goal of the present paper is the determination of precise atmospheric parameters, i.e. the effective temperature, $\Teff$, the surface gravity, log~g, the iron abundance, [Fe/H], and the microturbulence velocity, $\vt$, for the selected sample of, presumably, dwarf stars. The next papers in the series will concern with calculations of the NLTE abundances for many chemical elements from Li to Eu and analysis of the obtained abundance trends. 
Taking advantage of the Shane/Hamilton and CFHT/ESPaDOnS high-resolution 
 observational material with sufficient spectral coverage 
and the NLTE line formation for Fe~I-Fe~II, we determined stellar atmosphere parameters spectroscopically from lines of iron in the two ionization stages, Fe~I and Fe~II. 

The spectroscopic methods are, in particular, useful for evaluating $\Teff$ and log~g of the VMP stars that are mostly distant, and, therefore, their temperatures cannot be reliably derived from photometric colours due to the uncertainty in the interstellar reddening data and, at present,  surface gravities cannot be calculated from the trigonometric parallax because it is either rather uncertain or non-measurable.
 We look forward to seeing soon accurate parallaxes from Gaia\footnote{{\tt http://sci.esa.int/gaia/}}.
 In the literature there is no consensus on a validity of the excitation temperatures, $T_{\rm exc}$, derived through forcing no dependence of the Fe~I line abundance on the excitation energy of the lower level, $\eexc$.
For their sample of VMP cool giants \citet{Cayrel2004} reached in LTE good agreement between the photometric and spectroscopic temperatures  
after they excluded strong lines with $\eexc$ = 0~eV. 
A similar approach (LTE, but not cutting $\eexc$ = 0~eV) yielded the same outcome for the VMP near-main-sequence stars in \citet{2013ApJ...769...57F}, while up to several hundred degrees cooler excitation temperatures than the photometric ones were found for the cool giants. 

\citet{Ruchti2013} determined surface gravities of the stellar sample selected from the RAVE survey, using a grid of the NLTE abundance corrections from \citet{lind2012}. For common $\Teff$ in the NLTE and LTE analysis, they found systematic biases in log~g of up to 0.2~dex in the metallicity range $-2 <$ [Fe/H] $<-0.5$ and up to 0.3~dex for the more metal-poor stars.

The NLTE effects for lines of Fe~I were accounted for by \citet{Bensby2014}
in their study of 714 F and G dwarf stars, 
 by applying the NLTE abundance corrections interpolated in the grid of \citet{lind2012}. They found minor shifts between NLTE and LTE, with $\Delta\Teff$ = $-12\pm28$~K, $\Delta$log~g = +0.012 $\pm$ 0.059, and $\Delta\Eps({\rm Fe}) = -0.013\pm0.016$, on average. This is because 
the majority of their stars are close-to-solar metallicity and mildly metal-deficient ones, with [Fe/H] $> -1.2$, and the NLTE abundance corrections are small, at the level of few units of hundredth, as predicted by \citet{mash_fe} and \citet{lind2012} for this metallicity range. Furthermore, the NLTE calculations of \citet{lind2012} resulted in small departures from LTE for Fe~I until the extremely low metallicities because they were performed assuming a high efficiency of 
the Fe+H collisions and using the Drawinian rates \citep{Drawin1968, Drawin1969}.
For nearby stars with very good {\sc Hipparcos} parallaxes ($\Delta\pi/\pi < 0.05$), \citet{Bensby2014} found that essentially all stars with log~g $>$ 4.2 and $\Teff <$ 5650~K do not show ionization equilibrium between Fe~I and Fe~II, when determining the surface gravity from {\sc Hipparcos} parallaxes. 
\citet{Bensby2014} suggested that classical plane-parallel (1D) models have limitations and cannot properly handle excitation balance and/or ionization balance, and they applied, therefore, empirical corrections to the atmospheric parameters from ionization balance.  In this study we check with our stellar sample, whether spectroscopic methods of $\Teff$ and log~g determination from lines of Fe~I and Fe~II have any limitations and what they are.

The paper is structured as follows. In Sect.~\ref{Sect:observations} we describe the stellar sample, observations, and their reduction. Kinematical properties of the selected stars are calculated in Sect.~\ref{Sect:Kinematics}. They are used to identify a membership of individual stars to the galactic stellar populations. Section~\ref{Sect:method} concerns with the methodical issues. Stellar effective temperatures, surface gravities, iron abundances, microturbulence velocities, and masses are derived in Sect.~\ref{Sect:parameters}. The obtained results are discussed in Sect.~\ref{Sect:comparisons}. 
Our conclusions are presented in Sect.~\ref{Sect:Conclusions}.

\section{Stellar sample, observations, spectra reduction}\label{Sect:observations}

The stars were selected from the [Fe/H] catalogue of \citet{cayrel2001} based on the following criteria. 
\begin{enumerate}
\item The stars have a declination of $\delta > -20^\circ$ to be observed at the nothern sky.
\item The selected stars should cover as broad as possible metallicity range and be uniformly distributed, with 2-3 stars in each 0.1~dex metallicity interval.
\item The stellar sample should be homogeneous in temperature and luminosity. We selected the F-G-K dwarfs and subgiants based on the literature data.
\item Binaries, variables, and stars with any chemical peculiarity (carbon-enhanced stars, low [$\alpha$/Fe] stars, etc.) were excluded.
\item For testing purposes the stellar sample should include a dozen of well-studied stars of various metallicities, for which their stellar atmosphere parameters $\Teff$ and log~g were determined in the literature, presumably, from the non-spectroscopic methods. 
\end{enumerate}

Thus,  50 stars, in total, were selected, and they cover the $-2.6 <$ [Fe/H] $< +0.3$ metallicity range.

Spectra of  48 stars were obtained using the Hamilton Echelle Spectrograph mounted on the Shane 3-m telescope of the Lick observatory during the two observation runs March 15-16, 2011 and January 5-11, 2012. Most stars were observed, at least, twice. The resolving power is $R = \lambda/\Delta\lambda \simeq$ 60\,000, and the spectral coverage is 3700~\AA\ to 9300~\AA. The signal-to-noise ratio ($S/N$) at 5500~\AA\ is higher than 100 for most stars. 
The exception is BD\,$-04^\circ$~3208, where $S/N$($\lambda$ = 5500\AA) $\simeq$ 70. It is worth noting that the spectra suffer from the fringing effect in the infrared band.

For  two stars, we used the spectra observed with the ESPaDOnS echelle spectrograph at the 3.6-m telescope of the Canada-France-Hawaii (CFH) observatory in queued service observing mode during several nights in 2011 and 2012. The spectra cover 3700~\AA\ to 10\,450~\AA. 
They have $R \simeq$ 81\,000 and $S/N \simeq 150$ at 5030~\AA\ and higher value of about 200 at 8090~\AA. 

The Shane/Hamilton spectra were reduced with the IRAF package following standard procedures that include the aperture definition, flat-field correction, background subtraction, one-dimensional spectra extraction, wavelength calibration, and continuum normalization. The wavelength calibration was performed using the lamp with the titanium cathode in argon environment. Such a lamp is rare in the astronomical observations, and an appropriate list of the Ti and Ar lines was compiled by \citet{pakhomov2013}. The uncertainty in wavelength calibration is estimated to be 0.006~\AA.

The CFHT/ESPaDOnS spectra were reduced with the special reduction package 
 Libre-ESpRIT\footnote{{\tt http://www.cfht.hawaii.edu/Instruments/Spectroscopy/Espadons}}, which includes two steps. The first one performs a geometrical analysis from a sequence of calibration exposures, and the second step achieves spectrum optimal extraction in itself, using the geometrical information derived in the first step.
The reduction procedure returns the normalised spectra.

To improve the statistics of the VMP stars, we include the well-studied halo star HD\,140283 in our sample. Its high-quality observed spectrum was taken from the {\sc ESO UVESPOP} survey \citep{2003Msngr.114...10B}. For seven stars our observational material was complemented with the data from different sources. We employed the spectra obtained by Klaus Fuhrmann with the fiber optics Cassegrain
echelle spectrograph FOCES at the 2.2-m telescope of the Calar
Alto Observatory in 1996 to 2000 \citep{Fuhrmann1998,Fuhrmann2004}. The star BD$-04^\circ$\,3208 was observed using UVES/VLT in April 2001 within our project 67.D-0086A (PI: T. Gehren). Details of spectra reduction were described by \citet{Mashonkina2003}. For HD\,49933 its high-quality spectrum was observed with the HARPS spectrograph at the 3.6-m ESO La Silla telescope. Details of spectra reduction were described by \citet{Ryabchik2009}.

This paper deals with 51 stars, in total. 
The investigated stars together with characteristics of the observed spectra are listed in Table\,\ref{Tab:observations}. 

\section{Kinematic properties of the selected stars}\label{Sect:Kinematics}

The stellar kinematics is closely related to stellar populations in the Galaxy, and it is commonly applied to identify a membership of given star to the galactic stellar populations, namely the thin disk, the thick disk, and the halo.

The galactic space velocity components U, V, W were calculated using the equations and formalism of \citet{1987AJ.....93..864J}. They were defined with respect to the local standard of rest (LSR), adopting the standard solar motion (U, V, W) = (10.00, 5.25, 7.17)\,\kms\ of \citet{1998MNRAS.298..387D}. In computations of the X, Y, Z-coordinates, we used the best current estimate of the Galactocentric distance of the Sun $R_G$ = 8.0\,kpc, which was inferred from a comparison of different statistical techniques \citep{2013ARep...57..128M}.
The parallaxes and proper motions were taken from the updated version of the {\sc Hipparcos} catalogue \citep{Hipp2007A&A...474..653V} and the radial velocities from the {\sc Hipparcos} Input catalogue \citep{1993BICDS..43....5T}. The obtained results are presented in Table\,\ref{kin}. The star BD\,$-13^\circ$\,3442 is missing in the {\sc Hipparcos} catalogue, and it was assigned to the galactic halo based on its very low Fe abundance, with [Fe/H] = $-2.62$ (Sect.\,\ref{Sect:parameters}), and high proper motion as indicated by the Simbad database\footnote{{\tt http://simbad.u-strasbg.fr/simbad/}}. 

\begin{figure}
\epsscale{1.0}
\plotone{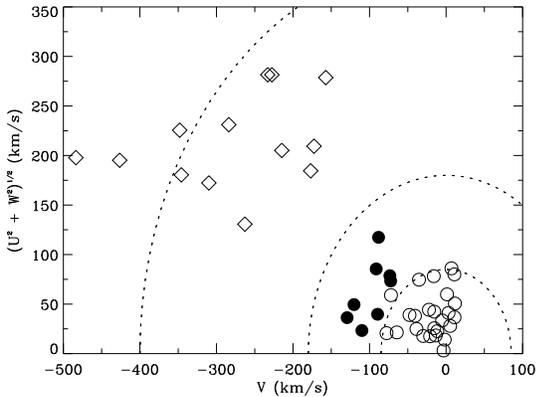}
\caption{Toomre diagram for the investigated stars from the perspective of the local standard of rest. Dashed curves delineate constant peculiar space velocities $v_{pec} = (U^2+V^2+W^2)^{1/2} = $~85, 180, and 400\,\kms. Different symbols show the thin disk (open circles), thick disk (filled circles), and halo (open rhombi) stars. \label{Fig:toomre}}
\end{figure}

Following \citet{Fuhrmann2000} and later studies \citep[for example][]{2004A&A...425..697C,Bensby2010A&A...516L..13B}, we identify the stars with peculiar space velocities with respect to the LSR in the range 85~\kms\ $< v_{pec} <$ 180~\kms\ as belonging to the thick disk stellar population, the stars with $v_{pec} <$ 85~\kms\ to the thin disk, and all the halo stars of our sample have $v_{pec} >$ 200~\kms\ (Fig.\,\ref{Fig:toomre}). The star G090-003 is not shown in Fig.\,\ref{Fig:toomre}, because its peculiar space velocity, $v_{pec}$ = 847 $\pm$ 1580~\kms, is rather uncertain due to large  error of the {\sc Hipparcos} parallax, $\pi$ = 1.12 $\pm$ 2.14~mas. For six stars with peculiar space velocities close to $v_{pec}$ = 85~\kms, their membership to either the thin disk or the thick disk cannot be decided using only the kinematic properties. Following \citet{Fuhrmann1998}, we also took into account their chemical properties, as determined in Sect.\,\ref{Sect:parameters}. For example, HD\,30562, with $v_{pec}$ = 93~\kms, was assigned to the thin disk because it has a supersolar Fe abundance, with [Fe/H] = 0.17, and it does not reveal Mg enhancement relative to Fe.

\section{Method of calculations}\label{Sect:method}

Stellar parameters of the selected sample were determined in this study applying different methods, including the spectroscopic one based on the NLTE analysis of lines of Fe~I and Fe~II.
This section describes calculations of the theoretical spectra.

\subsection{Codes and model atmospheres}

The statistical equilibrium (SE) of Fe~I-Fe~II was calculated with a comprehensive model atom treated by \citet{mash_fe}.  The main source of the uncertainty in the NLTE results for Fe~I is poorly known inelastic collisions with hydrogen atoms. We employed the formula of \citet{Drawin1968, Drawin1969}, as implemented by \citet{Steenbock1984}, for allowed $b-b$ and $b-f$ transitions and a simple relation between hydrogen and electron collisional rates, $C_H = C_e \sqrt{(m_e/m_H)} N_H/N_e$, for forbidden transitions, following \citet{1994PASJ...46...53T}. \citet{mash_fe} recommended to scale the Drawin rates by a factor of \kH\ = 0.1 based on their analysis of element abundances derived from the two ionization stages, Fe~I and  Fe~II, in the selected VMP stars. \citet{Bergemann_fe_nlte} estimated a larger value of \kH\ = 1. 
In this study, \kH\ was further constrained using the benchmark stars.

The coupled radiative transfer and SE equations were solved with a revised version of the DETAIL code \citep{detail}. The update was presented by \citet{mash_fe}. The obtained departure coefficients were then used by the codes binmag3  \citep{oleg}, synthV-NLTE  \citep[Vadim Tsymbal, private communication, based on SynthV code][]{synthv} and SIU \citep{Reetz} to calculate the synthetic line profiles. 

 We used the MARCS model structures \citep{Gustafssonetal:2008}\footnote{\tt http://marcs.astro.uu.se}, which were interpolated for given \teff, log~g, and [Fe/H] using a FORTRAN-based routine written by Thomas Masseron\footnote{{\tt http://marcs.astro.uu.se/software.php}}.

\subsection{Selection of iron lines}

The investigated lines were selected according to the following criteria.

(i) The lines should be almost free of visible/known blends in the solar spectrum.

(ii) For Fe~I the linelist should cover a range of excitation energies of the lower level, $\eexc$, as large as possible to investigate
the excitation equilibrium of neutral iron. On the other hand, the low-excitation ($\eexc <$ 1.2~eV) lines in metal-poor stars often appear to have higher abundances than lines with higher excitation energy \citep[see, for example][]{Cayrel2004,Lai2008}. The calculations with ab initio 3D, time-dependent, hydrodynamical model atmospheres of cool stars predict that the 3D effects for lines of Fe~I are strongly $\eexc$ dependent \citep{collet2007,2013A&A...559A.102D}. 
For example, the 3D abundance corrections amount to $-0.8$~dex and $-0.25$~dex for the $E_{\rm exc} = 0$ lines of Fe~I in the models \teff~/~log~g~/~[M/H] = 5860/4/$-2$ and 5850/4/$-1$, respectively, while they are $-0.2$~dex and $-0.07$~dex for the $E_{\rm exc} = 2$~eV lines and close to 0 for the $E_{\rm exc} = 4$~eV lines \citep{2013A&A...559A.102D}. To minimise a possible influence of the 3D effects on \teff\ determinations, we did not consider lines of Fe~I with $E_{\rm exc} < 2$~eV.

(iii) For both close-to-solar metallicity and VMP stars, lines of various strength have to be present in the spectrum to evaluate stellar microturbulence velocity $\vt$.

For individual stars, we also avoided using of saturated lines with the equivalent width $EW >$ 180\,m\AA\ to minimise an influence of possible uncertainty in the van der Waals damping constants on final abundances. 

The selected lines are listed in Table\,\ref{lines} along with their atomic parameters. For neutral iron we employed experimental $gf$-values from 
\citet{1979MNRAS.186..633B, 1982MNRAS.199...43B, 1982MNRAS.201..595B, 1991JOSAB...8.1185O}, and \citet{1991A&A...248..315B}. Two sets of oscillator strengths from \citet{mb2009} and \citet{ru1998} were inspected for Fe~II. Van der Waals broadening of the selected lines is accounted for using the most accurate data as provided by \citet{BPM} for Fe~I and \citet{2005A&A...435..373B} for Fe~II.

\subsection{Sun as a reference star}

To minimise the effect of the uncertainty in $gf$-values on the final
results, we applied a line-by-line differential NLTE and
LTE approach, in the sense that stellar line abundances were
compared with individual abundances of their solar counterparts.
The solar flux observations were
taken from the Kitt Peak Solar Flux Atlas \citep{Atlas}. The
calculations were performed with the MARCS model atmosphere $5777/ 4.44/ 0$. A microturbulence velocity of 0.9\,\kms\ was adopted. 

The solar LTE and NLTE abundances from individual lines are presented in Table\,\ref{lines}.  The NLTE calculatoins were performed using \kh\ = 0.5 (see sect.\,\ref{Sect:parameters}). For the Fe~I lines, the solar mean LTE and NLTE abundances are log~$A_{\rm LTE} = -4.57 \pm 0.09$ and log~$A_{\rm NLTE} = -4.56 \pm 0.09$. Hereafter, $A_X = N_X/N_{tot}$ is the element abundance taken relative to the total number of atoms. For lines of Fe~II, the NLTE abundance corrections do not exceed 0.01~dex, in absolute value, and the 
solar mean abundance amounts to 
$\eps{} = -4.56 \pm 0.05$ and $\eps{} = -4.50 \pm 0.05$, when using $gf$-values from \citet{mb2009} and \citet{ru1998}, respectively. Hereafter, the statistical abundance error is
the dispersion in the single line measurements about the mean:
$\sigma = \sqrt{\Sigma (x - x_i )^2 /(N - 1)}$, where N is the total number of lines used, x is their mean abundance, $x_i$ the individual abundance of each line. It is worth noting that an uncertainty of 0.06~dex in abundance from the Fe~II lines leads to an 0.12~dex uncertainty in log~g for solar-type stars. This is crucial for surface gravity determinations from the Fe~I/Fe~II ionization equilibrium. 

\section{Determination of stellar parameters}\label{Sect:parameters}

To derive atmospheric parameters of our stellar sample, the following strategy was applied. First, we selected the stars, for which their effective temperatures and surface gravities were determined reliably using the non-spectroscopic methods, namely $\Teff$ from the infrared flux method (IRFM) and log~g from the well-known relation between log~g, stellar mass $M$, $\Teff$, and absolute bolometric magnitude ${\rm M_{bol}}$:

\begin{equation}
[g] = [M] + 4[\Teff] + 0.4 ({\rm M_{bol}}-{\rm M_{bol\odot}}). \quad  \label{formula1}
\end{equation}
 
\noindent Here, square brackets denote the logarithmic ratio with respect to the solar
value. The ${\rm M_{bol}}$ magnitudes were obtained using the $V$ magnitudes from \citet{Olsen1983A&AS...54...55O,Olsen1993A&AS..102...89O}, the revised 
{\sc Hipparcos} parallaxes of \citet{Hipp2007A&A...474..653V}, and the bolometric corrections (BC) from \citet{Alonso1995}. For the Sun the absolute visual magnitude ${\rm M_{V\odot}} = 4.87$ and BC$_\odot = -0.12$ were adopted. The star's mass was estimated from its position in the ${\rm M_{bol}}$ - $\log\Teff$ diagram by interpolating in the   Dartmouth isochrones  \citep{Dotter2008}.
``Reliably'' means that 
multiple measurements of the IRFM temperature are available and the uncertainty in the {\sc Hipparcos} parallax based surface gravity, log~g$_{Hip}$, is less than 0.1~dex. This implies a parallax error of smaller than 10\,\%.
The 20 stars in our sample meet these requirements. They are listed in Table\,\ref{params} as the benchmark stars. 

Atmospheric parameters of the remaining stars were derived or improved spectroscopically from the NLTE analysis of lines of Fe~I and Fe~II. 

In the stellar parameter range, with which we concern, lines of Fe~I are weakened towards higher \teff, resulting in higher derived element abundance, while they are nearly insensitive to variation in log~g. In contrast, lines of Fe~II are only weakly sensitive to variation in \teff, while they are weakened with log~g increasing. If \teff\ is fixed, the surface gravity is obtained from the requirement that abundances from lines of Fe~I and Fe~II must be equal. One needs to be cautious, when deriving both \teff\ and log~g from Fe~I and Fe~II. Any shift in \teff\ leads to shifting log~g in the sense: an increase in \teff\ leads to an increase in log~g. Therefore, in this study we used non-spectroscopic (IRFM) determinations of \teff, where available, and for every star its final parameters, \teff/log~g/[Fe/H], were checked with the corresponding evolutionary track.

For each  benchmark star, abundances from lines of Fe~I and Fe~II were derived under various line-formation assumptions, i.e., NLTE with \kH\ = 0.1, 0.5, 1 and LTE, and we investigated which of them leads to consistent element abundances from both ionization stages. Our calculations showed that the departures from LTE are small for the [Fe/H] $> -1$ stars. Indeed, in this metallicity range the NLTE abundance correction, $\Delta_{\rm NLTE} = \log A_{\rm NLTE} - \log A_{\rm LTE}$, does not exceed 0.04~dex for \kH\ = 0.5, and the difference in NLTE abundances between \kH\ = 0.1 and 0.5 is smaller than 0.04~dex. The three of six [Fe/H] $< -1$ stars are cool, with $\Teff \le$ 5400~K, and they are not useful for constraining \kH\ because of small $\Delta_{\rm NLTE} \le$ 0.02~dex. 
 Table\,\ref{shtab} lists the abundance differences between Fe~I and Fe~II from the calculations with different \kH\ for the three benchmark stars with the largest NLTE effects. For every star, the acceptable abundance difference is achieved with \kH\ = 0.5 and 1. Hereafter, we choose \kH\ = 0.5 to perform the NLTE calculations for Fe I-FeII.

\begin{deluxetable}{rcccc}
\tabletypesize{\scriptsize}
\tablecaption{Abundance differences Fe I - Fe II in the selected three stars for different line formation scenarios. \label{shtab}}
\tablewidth{0pt}
\tablehead{
\colhead{}  & \colhead{LTE} &  \multicolumn{3}{c}{NLTE} \\
\colhead{HD}  & \colhead{} &  \colhead{\kH\ = 1} & \colhead{\kH\ = 0.5} & \colhead{\kH\ = 0.1} 
}
\startdata
  84937    & -0.06 & -0.02 &  0.00~ &  0.11~ \\
  94028    & -0.06 & -0.05 & -0.04~ & -0.01~   \\
 140283    & -0.06 & -0.04 & -0.02~ &  0.10~   
\enddata
\end{deluxetable}

Spectroscopic stellar parameters were obtained in the iterative procedure. For each star, our initial guess was $\Teff$ = $T_{\rm IRFM}$, if available, or $\Teff$ = $T_{b-y}$ and log~g = log~g$_{Hip}$.
Effective temperature and surface gravity were corrected until the ionisation equilibrium between Fe~I and Fe~II and the Fe~I excitation equilibrium were fulfilled.

\subsection{Effective temperatures}

\begin{figure}
\epsscale{1.0}
\plotone{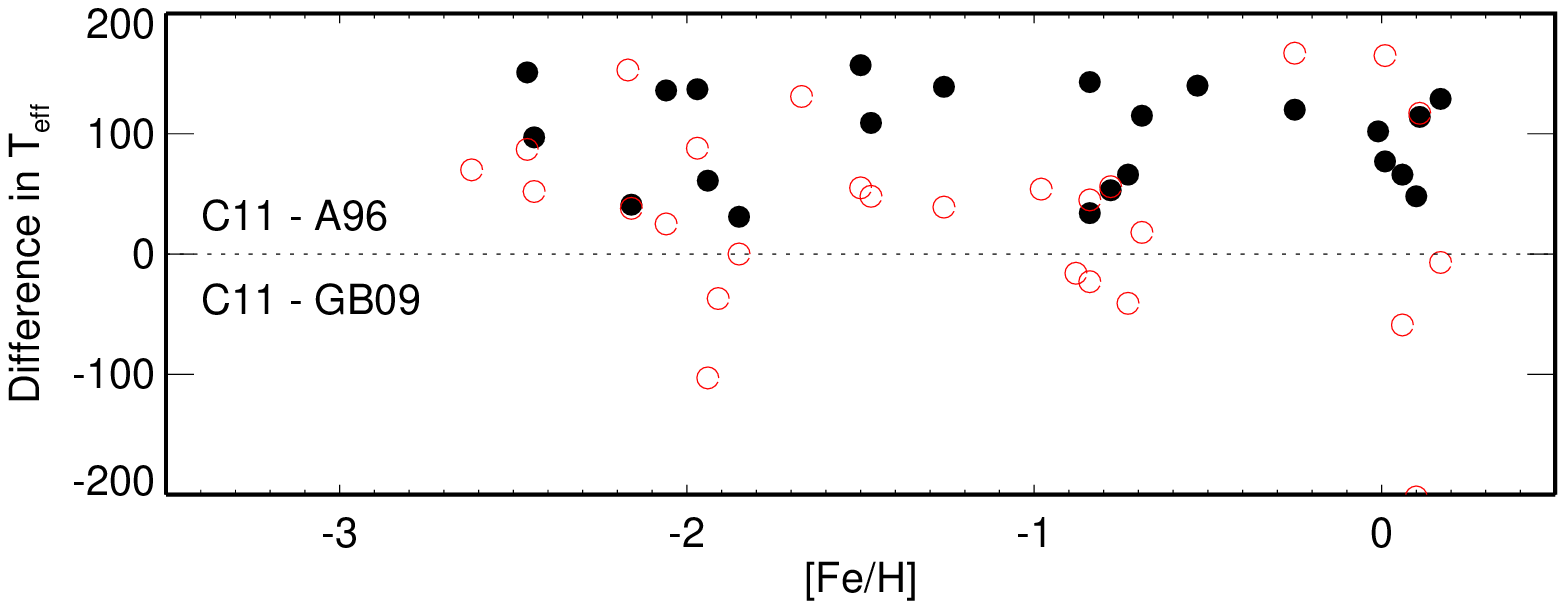}
\plotone{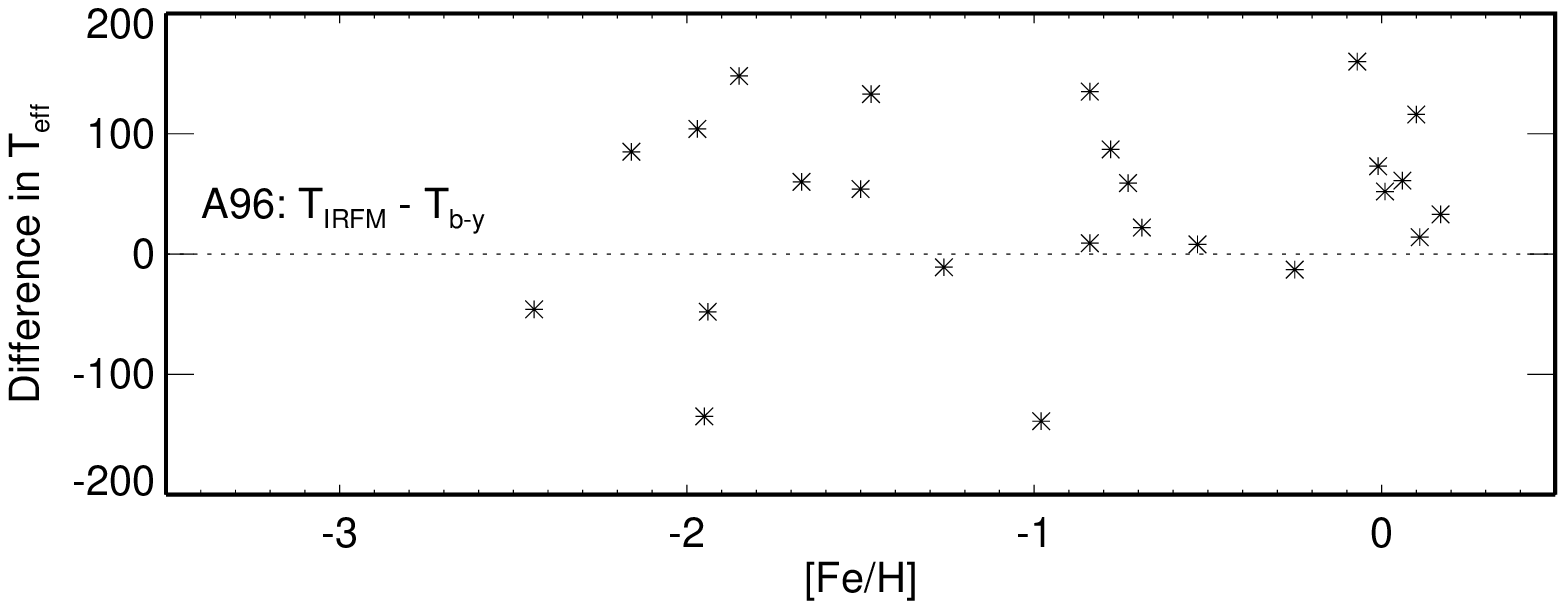}
\plotone{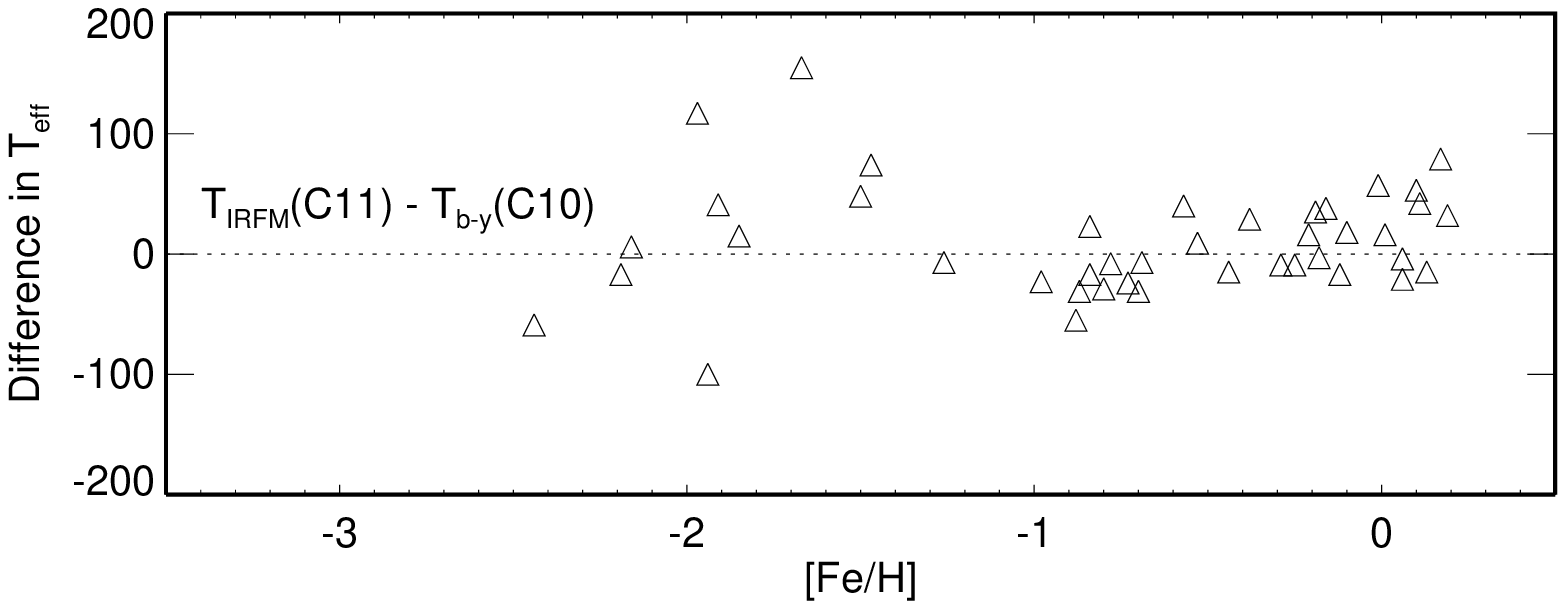}
\caption{Top panel: differences in the IRFM effective temperatures of the investigated stars between \citet[][C11]{Casagrande2011} and \citet[][A96]{Alonso1996irfm} (filled circles) and between C11 and \citet[][GB09]{GH2009A&A...497..497G} (open circles). Middle panel: differences $T_{\rm IRFM}$(A96) -- $T_{b-y}$\citep{Alonso1996}. Bottom panel: differences $T_{\rm IRFM}$(C11) -- $T_{b-y}$\citep[][C10]{Casagrande2010}. \label{Fig:teff_comp}}
\end{figure}

Although there are multiple measurements of the IRFM effective temperature for individual stars of our sample \citep{Alonso1996irfm,GH2009A&A...497..497G,Casagrande2010,Casagrande2011}, we found no study that provides the data for every investigated star. Figure\,\ref{Fig:teff_comp} (top panel) displays the differences between different sources for the stars in common with this work. It can be seen that the $\Teff$ scale of \citet[][C11]{Casagrande2011} is hotter than that of \citet[][A96]{Alonso1996irfm}, with $\Delta T_{\rm IRFM}$(C11 - A96) = 114 $\pm$ 62~K for 25 stars. 
The difference between C11 and \citet[][GB09]{GH2009A&A...497..497G} is smaller, $\Delta T_{\rm IRFM}$(C11 - GB09) = 32 $\pm$ 83~K for 26 stars. 
Here, we did not count the two outliers HD\,34411 and HD\,142373, with large uncertainties in $T_{\rm IRFM}$ of 479 K and 342 K, respectively, that were caused by using saturated 2MASS photometry.

We calculated four sets of the photometric temperatures from the $b-y$ and $V-K$ colour indices using two different calibrations of \citet[][C10]{Casagrande2010} and \citet{Alonso1996}. The $V$ magnitudes and $b-y$ colours were taken from \citet{Olsen1983A&AS...54...55O,Olsen1993A&AS..102...89O} for 45 stars, in total, and the $K$ magnitudes from the 2MASS catalogue \citep{2MASS2006AJ....131.1163S} for 24 stars, in total. 
Figure\,\ref{Fig:teff_comp} display the differences between $T_{\rm IRFM}$ and $T_{b-y}$ for both effective temperature scales.
We found that $T_{b-y}$ and $T_{V-K}$ agree well with those from the IRFM, when using the calibrations of C10, with  $T_{\rm IRFM}$(C11) -- $T_{b-y}$(C10) = 10 $\pm$ 46~K (42 stars) and $T_{\rm IRFM}$(C11) -- $T_{V-K}$(C10) = 22 $\pm$ 68~K (21 stars). 
For the \citet{Alonso1996} calibration, 
the differences are larger, with $T_{\rm IRFM}$(A96) -- $T_{b-y}$ = 41 $\pm$ 78~K (25 stars) and $T_{\rm IRFM}$(A96) -- $T_{V-K}$ = 72 $\pm$ 58~K (12 stars), although the statistics is small for the $V-K$ photometry.

For the {\it benchmark stars}, we aimed to have a homogeneous set of $\Teff$ based on a single source of the IRFM temperatures. The only study that provides $T_{\rm IRFM}$ for every benchmark star is GB09. Using their $T_{\rm IRFM}$, we could not achieve the Fe~I/Fe~II ionization equilibrium and the Fe~I excitation equilibrium
 for half benchmark stars, independent of applying either LTE or NLTE approach.
 No preference was, therefore, given to any source of $T_{\rm IRFM}$. Instead, for each star its temperature from A96, GB09, C10, and C11, where available, 
and the corresponding log~g$_{Hip}$ value were checked with
 the difference in NLTE abundances between Fe~I and Fe~II and a slope of the [Fe/H] versus $\eexc$ plot for Fe~I. For six stars, corrections up to 50~K were applied to the most appropiate $T_{\rm IRFM}$ to obtain the final effective temperature (Table\,\ref{params}). 


\subsection{Surface gravities, metallicities, and microturbulence velocities}

For each benchmark star, the NLTE abundances from the two ionization stages, Fe~I and Fe~II, were found to be consistent within 0.06~dex. Hereafter, 0.06~dex was considered as an admissible difference between Fe~I and Fe~II, when deriving spectroscopic stellar parameters of the remaining, non-benchmark, stars. 
For given star, we started from checking the photometric temperature, $T_{\rm IRFM}$ or $T_{b-y}$, and the {\sc Hipparcos} parallax based gravity, log~g$_{Hip}$. Effective temperature and surface gravity were allowed to vary within the corresponding error bars.
The procedure was iterated until the NLTE abundance difference Fe~I -- Fe~II is getting smaller than 0.06~dex, the excitation trend for Fe~I disappears, and lines of different equivalent width give consistent iron abundances.

\begin{figure}
\epsscale{1.0}
\plotone{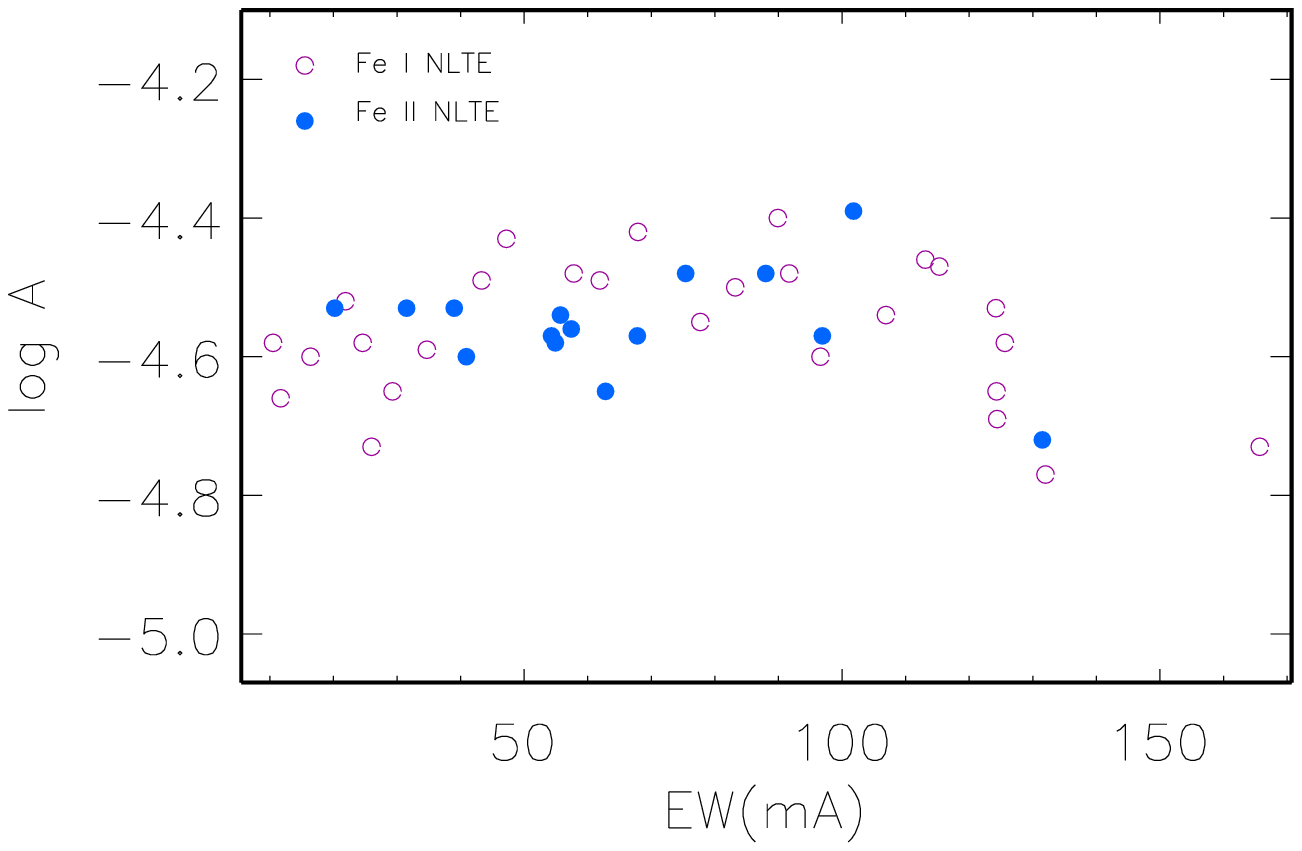}
\plotone{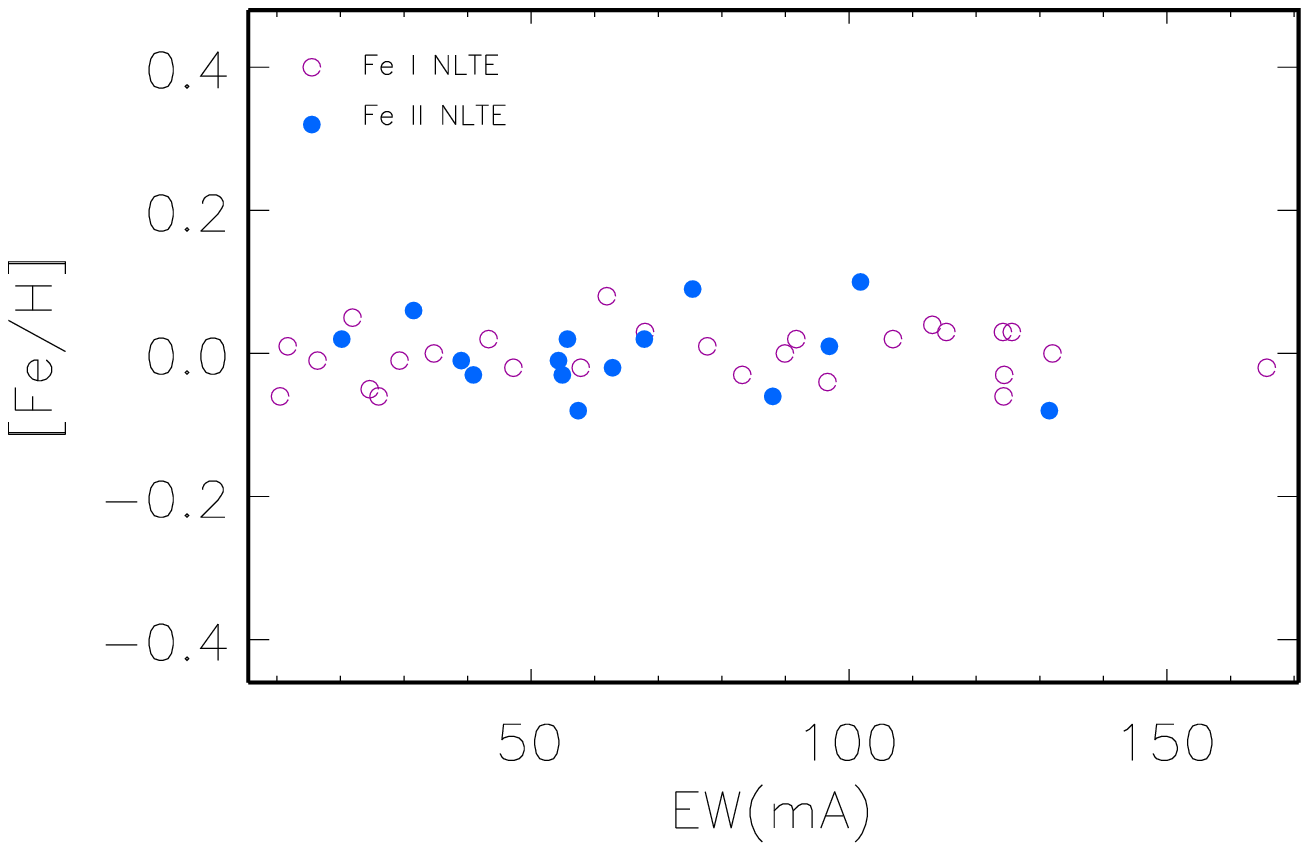}
\caption{Absolute (top panel) and differential (bottom panel) NLTE abundances from lines of Fe~I (open circles) and Fe~II (filled circles) in HD\,22484. \label{abs}}
\end{figure}

\begin{figure}
\epsscale{1.0}
\plotone{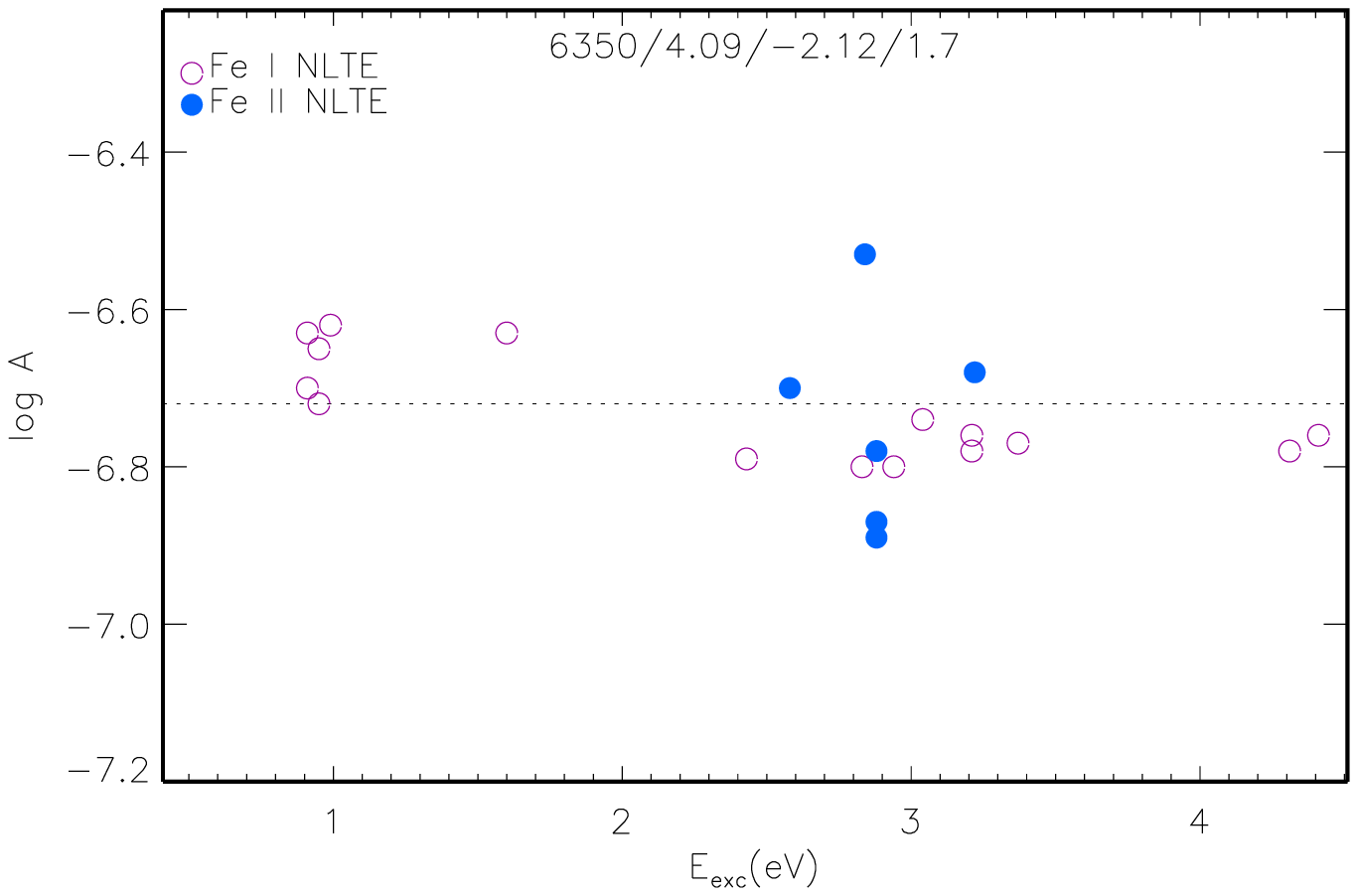}
\plotone{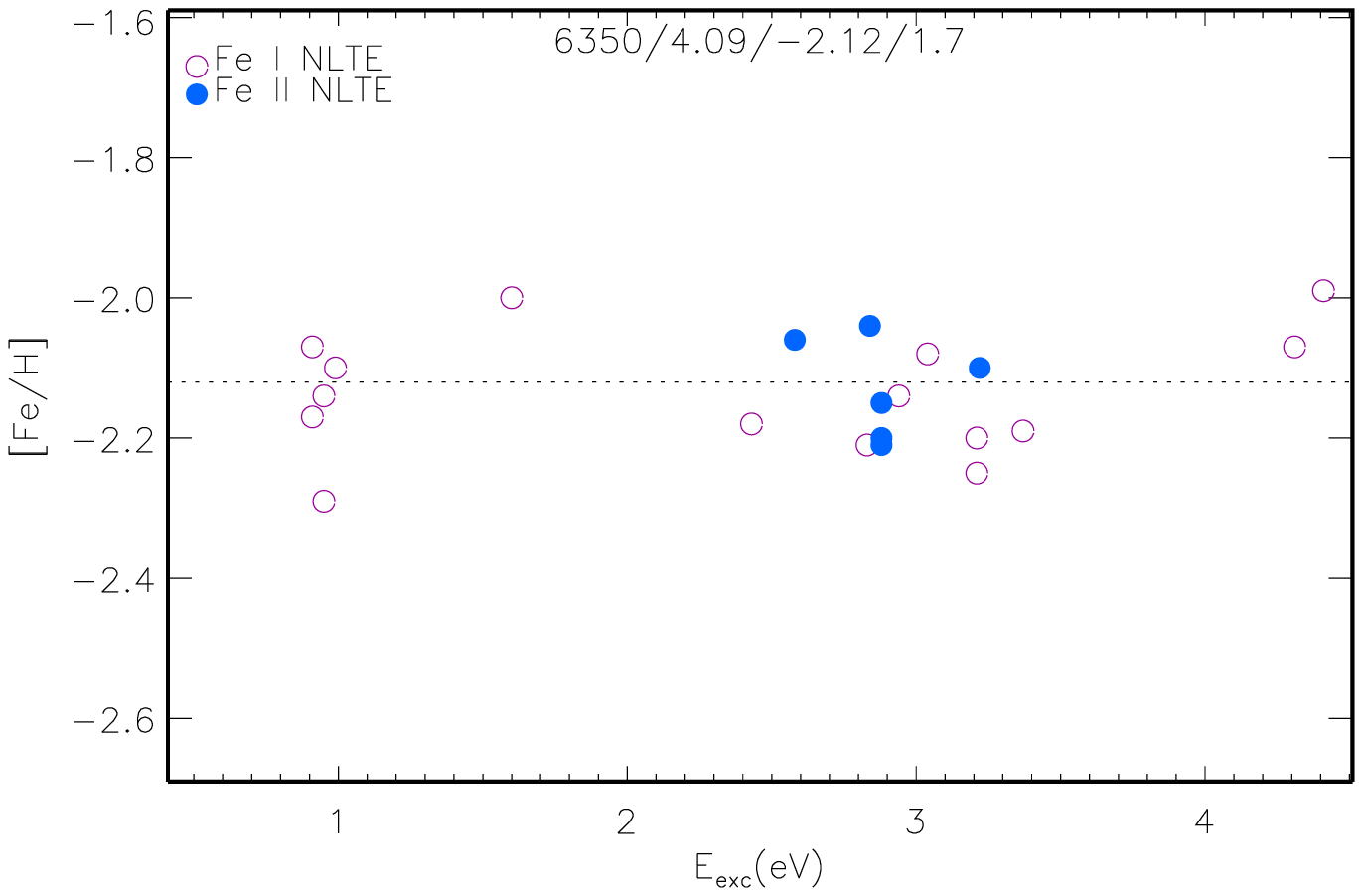}
\caption{Absolute (top panel) and differential (bottom panel) NLTE abundances from lines of Fe~I (open circles) and Fe~II (filled circles) in  HD~84937. \label{abs1}}
\end{figure}

\begin{figure}
\epsscale{.80}
\plotone{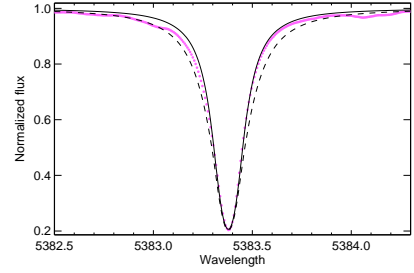}
\caption{Theoretical NLTE profiles of Fe~I 5383\,\AA\ computed with $\log C_{6} = -30.370$ \citep[][dashed curve]{BPM} and $\log C_{6} = -31.095$ \citet[][continuous curve]{K07} compared to the solar spectrum \citep[][bold dots]{Atlas}. Everywhere in the calculations, $\eps{Fe} = -4.54$. To fit the solar line profile with $\log C_{6} = -30.370$, one needs to reduce the iron abundance down to $\eps{Fe} = -4.72$.}\label{kb5383}
\end{figure}

We found that the differential approach largely removes a line-to-line scatter for the [Fe/H] $\ge -1.5$ stars. For example, when moving from the absolute to the differential NLTE abundances, the dispersion reduces from $\sigma$ = 0.10~dex to $\sigma$ = 0.04~dex for lines of Fe~I and from $\sigma$ = 0.08~dex to $\sigma$ = 0.05~dex for lines of Fe~II in HD\,22484 (6000/4.07/0.01, Fig.\,\ref{abs}). For more metal-poor stars, the differential approach is less efficient in removing a line-to-line scatter. 
For example, for HD~84937 (6350/4.09/$-2.12$) the scatter of data reduces for Fe~II, but not Fe~I lines (Fig.\,\ref{abs1}). When using the $\eexc >$ 2~eV lines of Fe~I, we obtained $\sigma = 0.035$~dex and 0.078~dex for the absolute and differential abundances, respectively.
 This is, probably, due to the uncertainties in the van der Waals damping constant, $C_{6}$. The lines, which can be measured in the VMP stars, have strong van der Waals broadened wings in the solar spectrum (for example Fe~I 5383\,\AA, Fig.\,\ref{kb5383}). 
For such a line, the uncertainty in the derived solar abundance is contributed from the uncertainties in both $gf$- and $C_{6}$-values. When dealing with the VMP stars, the differential analysis is able to cancel the uncertainty in $gf$, but not $C_{6}$.

Although our final results are based on using the $\eexc \ge$ 2~eV lines of Fe~I, 
 for most stars we also checked abundances determined from the lower excitation lines.  
We found no abundance difference between the low and high excitation lines for close-to-solar metallicity stars. For the most metal-poor stars, such as HD~84937, the absolute abundances reveal an excitation trend, but it disappears for the differential abundances (Fig.\,\ref{abs1}).

We did not determine stellar parameters under the LTE assumption and cannot evaluate the differences in log~g between using LTE and NLTE. However, the NLTE effects can be estimated based on our LTE and NLTE calculations for Fe~I-Fe~II and the sensitivity of iron lines to variation in log~g. Since the departures from LTE lead to weakened lines of Fe~I, but they do not affect lines of Fe~II until the extremely low metallicities, the ionization equilibrium Fe~I/Fe~II is achieved for higher gravities in NLTE than in LTE. The shifts in log~g increase towards lower metallicity and  surface gravity. For  our program stars they
 may be up to 0.1~dex at [Fe/H] $> -1.5$, up to 0.2~dex for $-2.2 <$ [Fe/H] $< -1.8$ and reach 0.45~dex for the most MP star of our sample BD\,$-13^\circ$\,3442, with [Fe/H] = $-2.62$.

\subsection{Checking atmospheric parameters with evolutionary tracks}

\begin{figure}
\epsscale{1.0}
\plotone{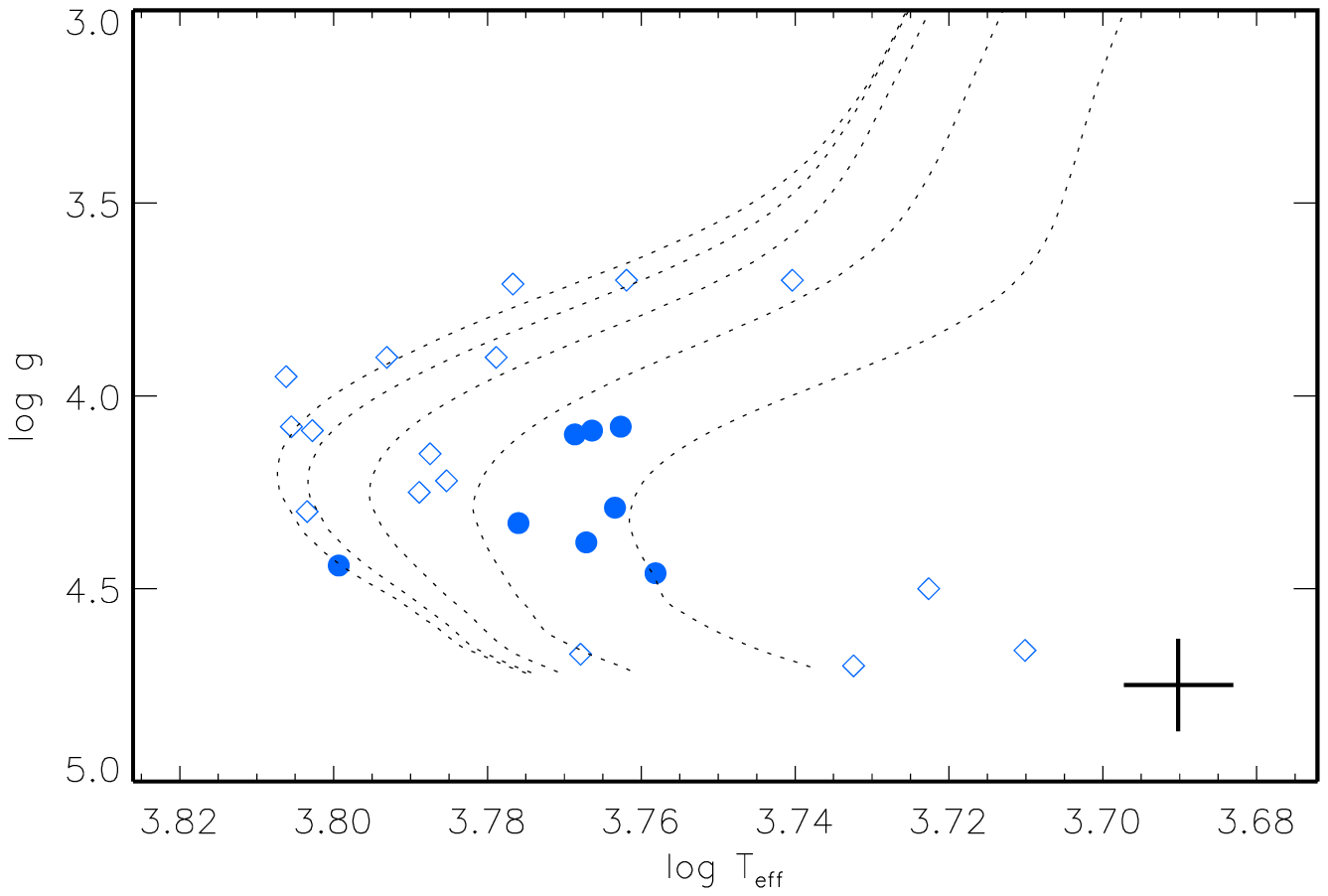}
\plotone{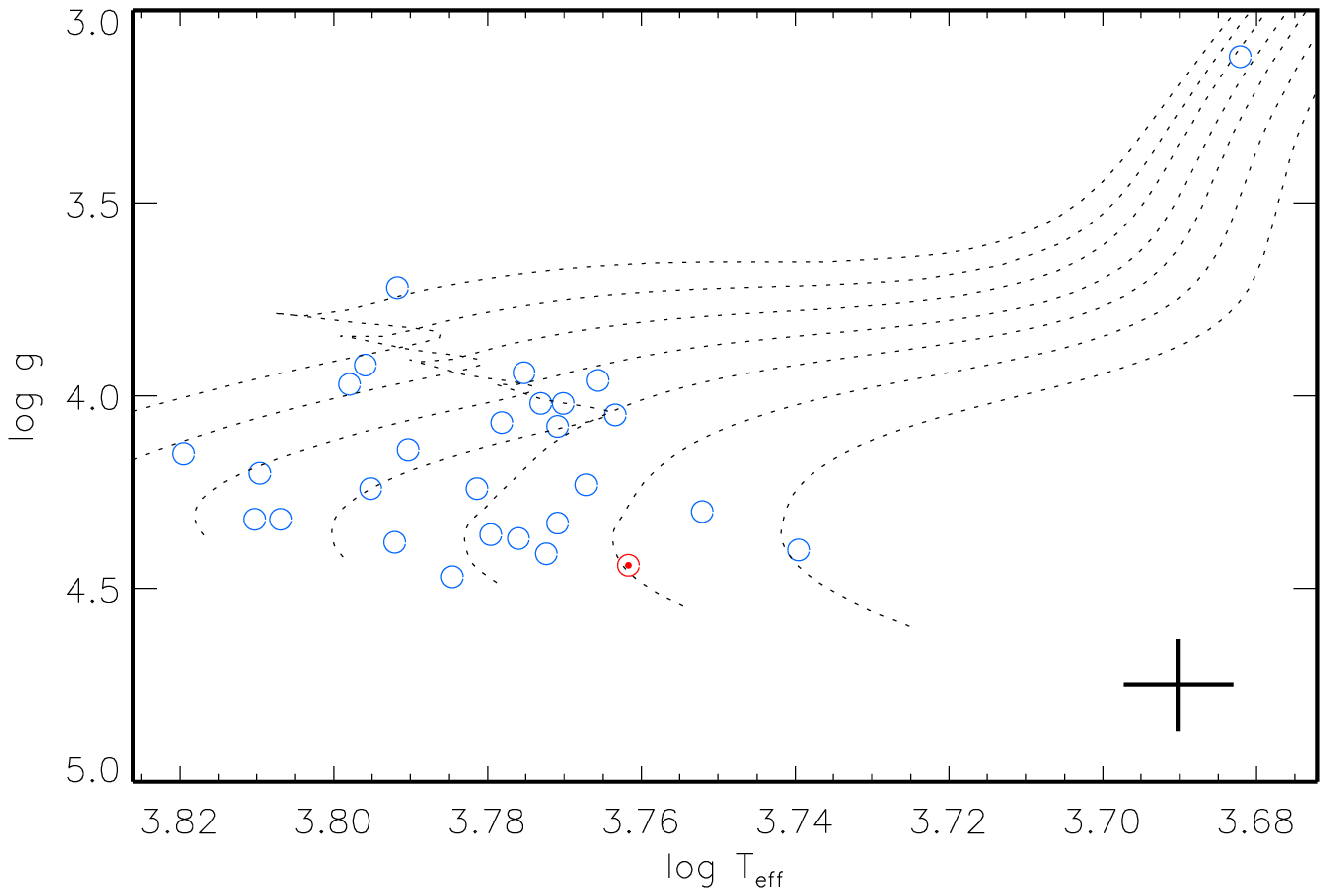}
\caption{Top panel: investigated thick disc (filled circles) and halo (diamonds) stars compared with the evolutionary tracks of $M$ = 0.75\,$M_\odot$ and [Fe/H] from $-2.75$ to $-0.75$ (from left to right), with a step of 0.5 dex. Bottom panel: the thin disc stars (open circles) compared with the evolutionary tracks of the solar metallicity and the stellar mass varying between $M$ = 0.9\,$M_\odot$ and 1.5\,$M_\odot$, with a step of 0.1$M_\odot$. The crosses in each panel indicate log~g and \teff\ error bars of 0.12~dex and 80~K, respectively.  \label{halo_thick}}
\end{figure}

For each star the obtained effective temperature and surface gravity were checked by comparing its position in the log~g -- \teff\ plane with the theoretical evolutionary track of given metallicity and $\alpha$-enhancement in the \citet{Yi2004} grid (Fig.\,\ref{halo_thick}). An $\alpha$-enhancement was assumed to be equal to [Mg/Fe] as derived in this study and presented in Table\,\ref{kin}. 
Stellar masses corresponding to the best fit evolutionary tracks are indicated in Table\,\ref{params}. 
They range between 0.60 and 0.85 solar mass for the halo and thick disk stars that does not contradict with their evolutionary status and old age. 
For the thin disk stars, their masses range between 0.85\,$M_\odot$ and 1.5\,$M_\odot$ and ages between 1~Gyr and 9~Gyr. 

For  two halo cool dwarfs we had to revise \teff~/~log~g, to  enforce the stars to sit on the main sequence of the appropriate evolutionary track.
We stress that the changes in atmospheric parameters were not allowed to destroy the ionization equilibrium between Fe~I and Fe~II and the Fe~I excitation equilibrium. 
For HD\,64090, we adopted \teff~ = 5400~K, which is close to $T_{\rm IRFM}$ = 5440~K (A96), and log~g = 4.70, which is 1.5$\sigma_{\rm log g}$ higher than log~g$_{Hip}$. 
For BD\,+66$^\circ$0268 our final temperature, \teff~ = 5300~K, is close to $T_{\rm IRFM}$ = 5280~K (A96) and log~g = 4.72 is 2$\sigma_{\rm log g}$ higher than log~g$_{Hip}$.

\subsection{Final atmospheric parameters}

\begin{figure}
\epsscale{1.0}
\plotone{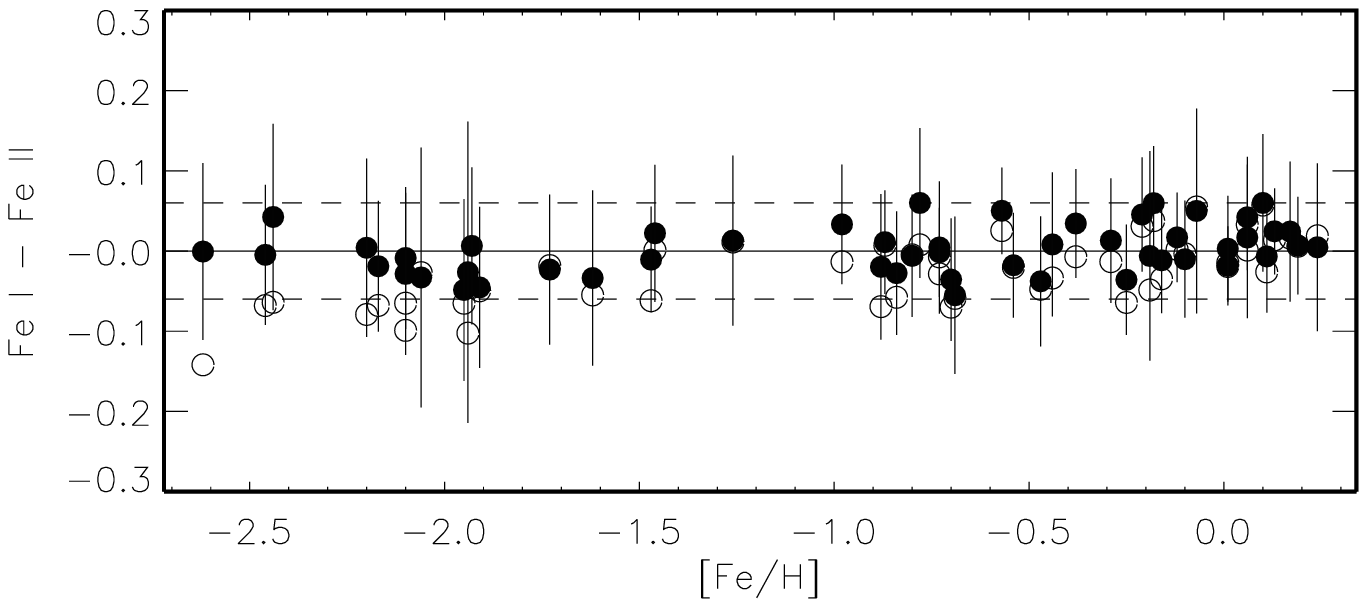}
\plotone{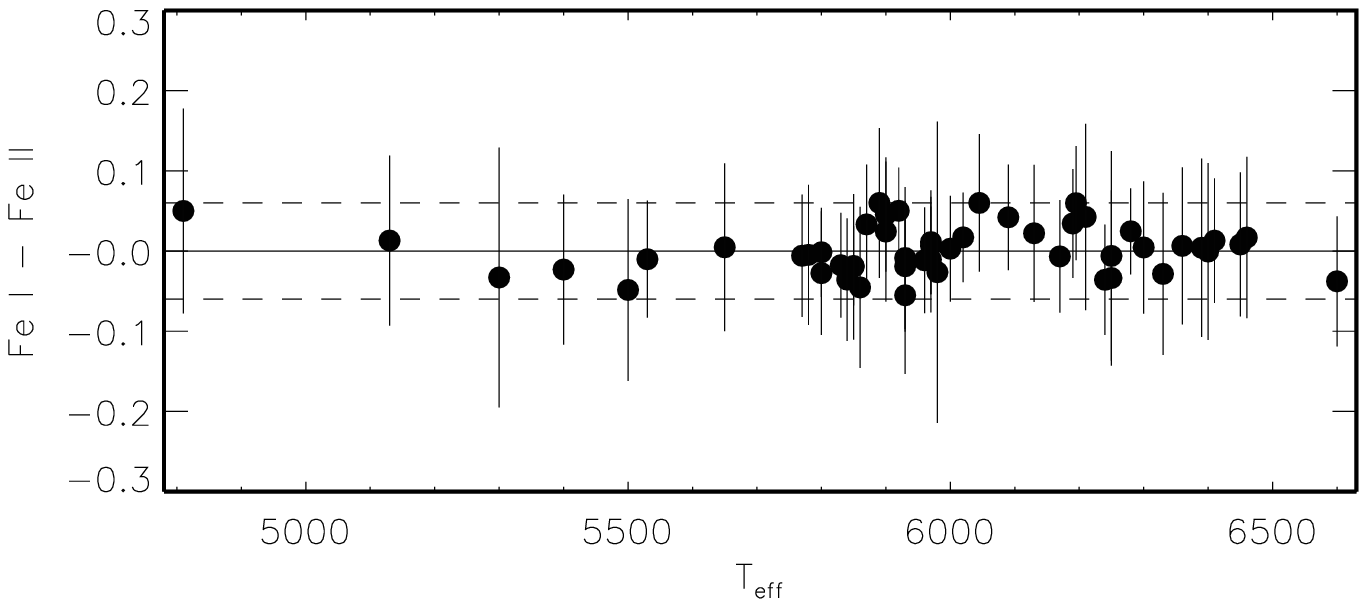}
\caption{NLTE (filled circles) and LTE (open circles, top panel only) abundance differences Fe~I -- Fe~II for the investigated stars. Dashed lines indicate an admissible difference of $\pm$0.06~dex. \label{tgf50}}
\end{figure}

The final atmospheric parameters together with the NLTE and LTE abundance differences between Fe~I and Fe~II are presented in Table\,\ref{params}. For the star's metallicity we adopted [Fe/H] determined from lines of Fe~II, although the difference in NLTE abundances between Fe~I and Fe~II nowhere exceeds 0.06~dex.
The Fe~I -- Fe~II abundance differences are also shown in Fig.~\ref{tgf50}. In NLTE, they reveal no trends with any atmospheric parameter, either [Fe/H], or $\Teff$, or log~g.

\begin{figure}
\epsscale{1.0}
\plotone{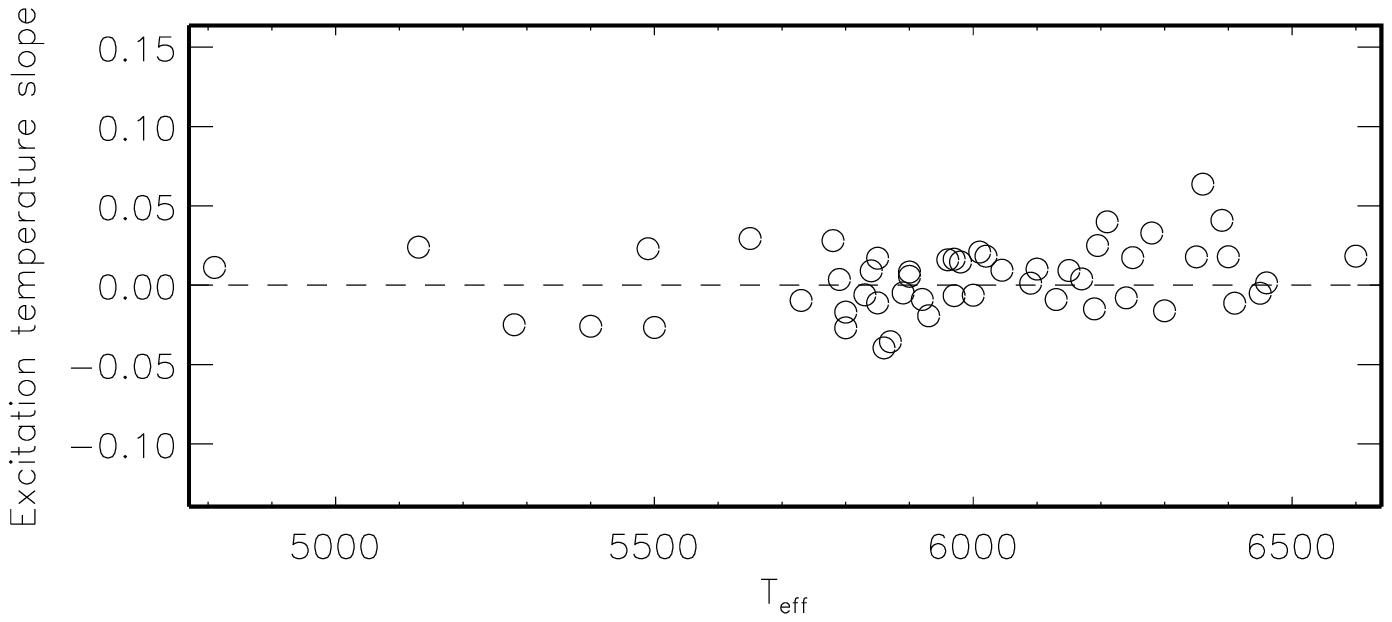}
\plotone{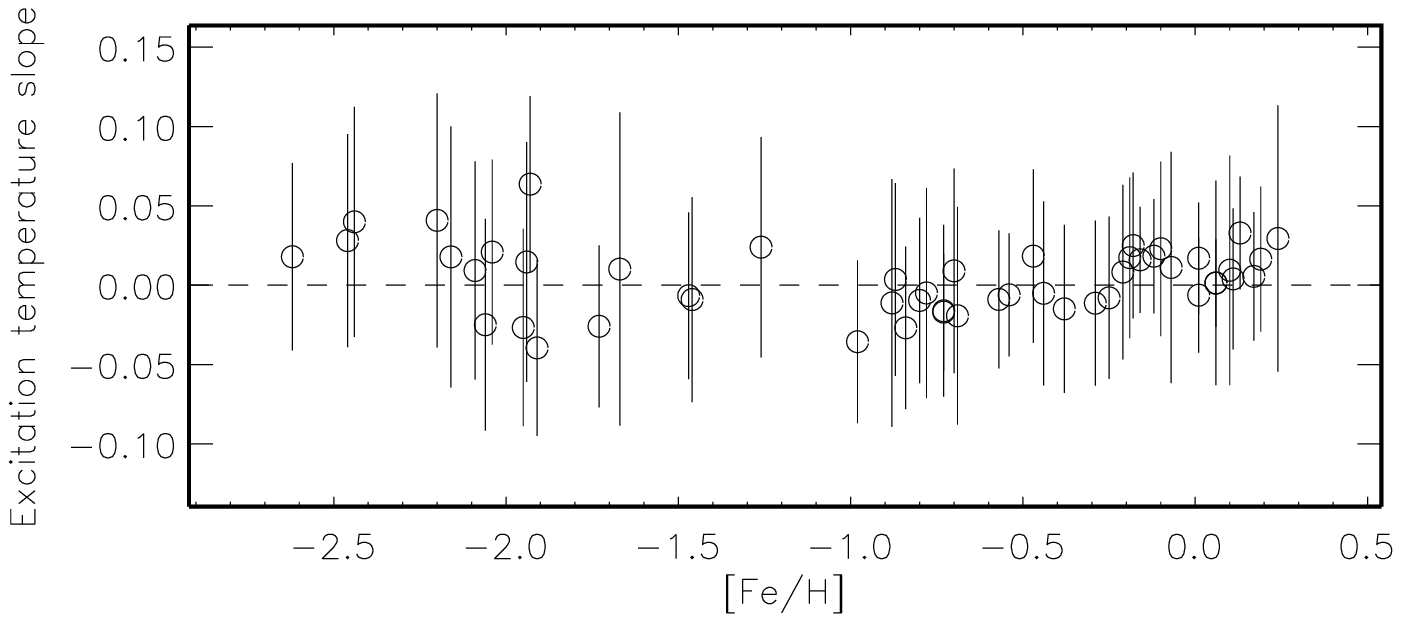}
\caption{Excitation temperature slopes from Fe~I lines for the investigated stars. \label{slopes}}
\end{figure}

Figure~\ref{slopes} displays the excitation temperature slopes, d[Fe/H]/d$\eexc$, calculated from lines of Fe~I in the individual stars. 
 The data reveal no trend with effective temperature, and $d$[Fe/H]/$d\eexc$ = 0.0044~dex/eV, on average. The excitation slope seems to depend slightly on metallicity at [Fe/H] $\ge -1$, although nowhere its magnitude exceeds the error bars. The scatter of data is larger for the [Fe/H] $< -1.8$ than less metal-poor stars due to smaller number of observed lines of Fe~I.

\begin{figure}
\epsscale{1.0}
\plotone{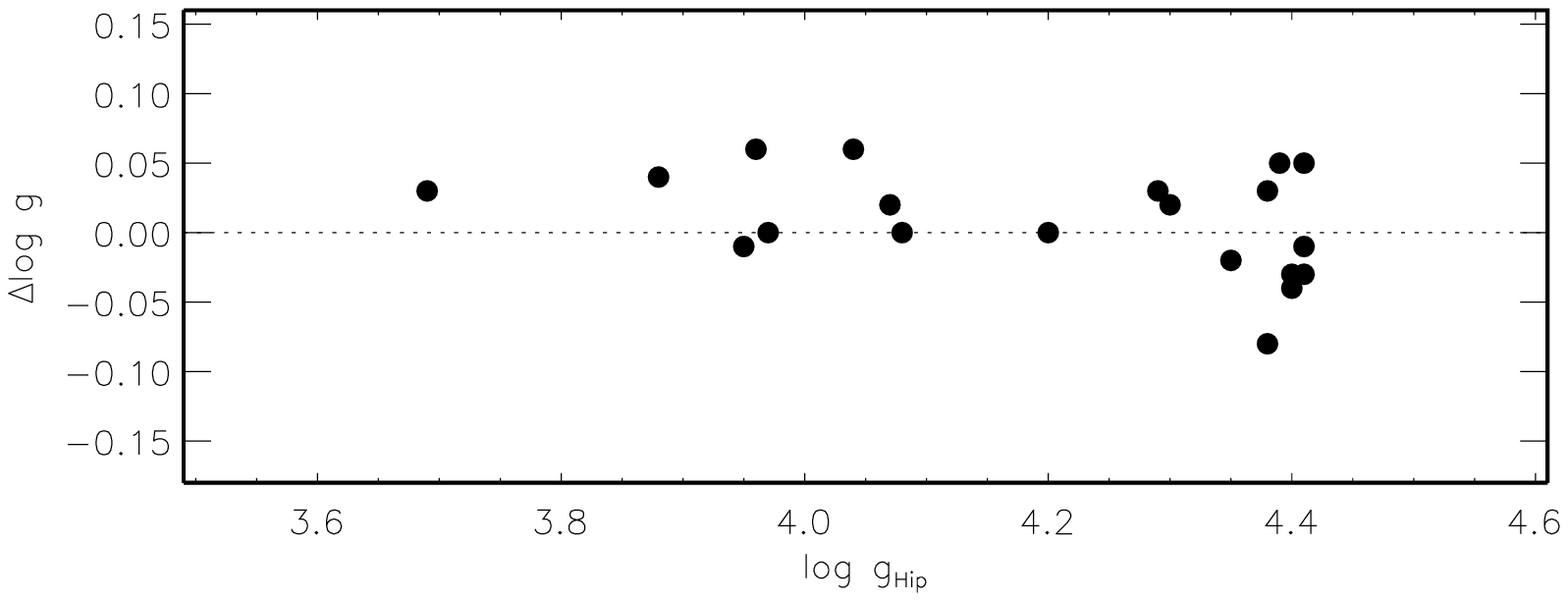}
\plotone{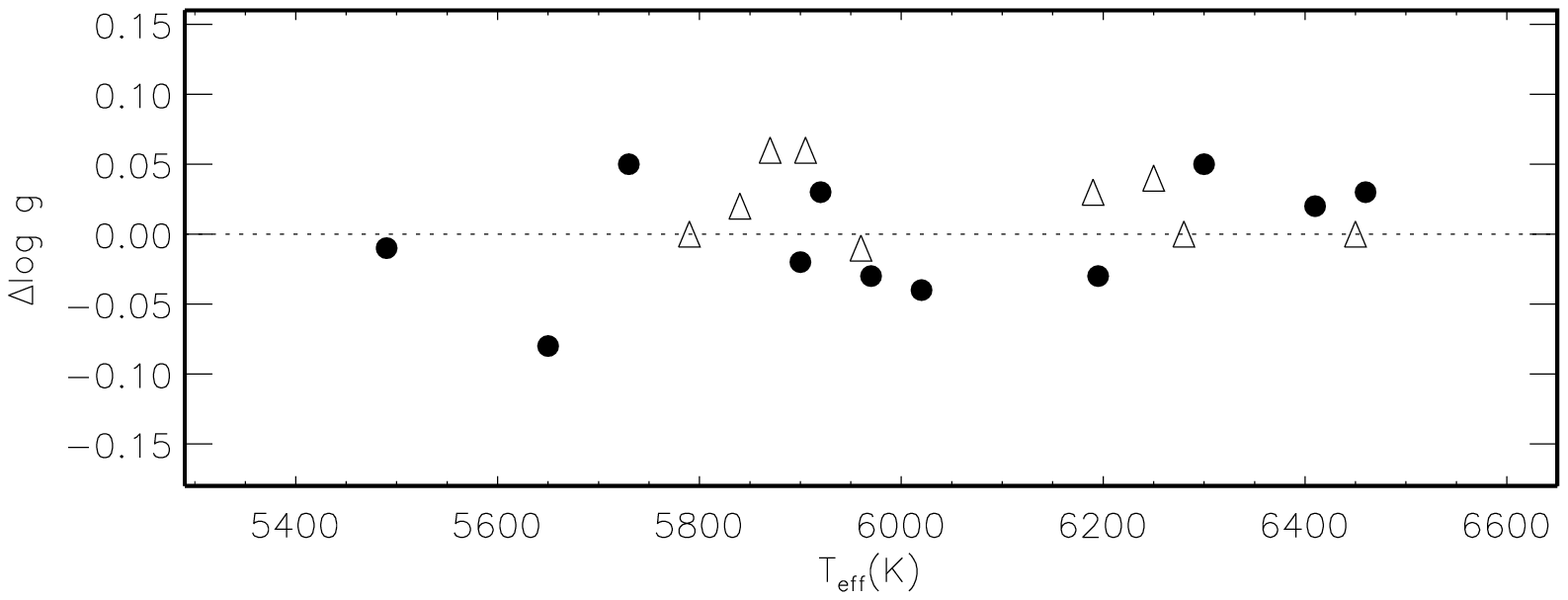}
\caption{Differences in surface gravity from the Fe~I/Fe~II NLTE ionization equilibrium and Hipparcos parallax methods for 20 stars with the spectroscopically derived atmospheric parameters and a relative parallax uncertainty less than 10\,\%. In the bottom panel, the filled circles and open triangles show stars with log~g$_{Hip} >$ 4.2 and $\le$ 4.2, respectively. 
 \label{sp_hipp}}
\end{figure}

In Fig.~\ref{sp_hipp} we compare surface gravities from the Fe~I/Fe~II NLTE ionization equilibrium and Hipparcos parallax methods for the 20 stars, which do not belong to the benchmark stellar sample, but have a relative parallax uncertainty less than 10\,\%. 
The differences are minor, with $\Delta$log~g(Sp -- Hip) = $0.008\pm0.037$~dex, on average, and they do not show any trends with surface gravity or effective temperature.
Having in mind that for each benchmark star the Fe~I/Fe~II ionization equilibrium was achieved with log~g from the Hipparcos parallax method, we infer that the spectroscopic method of gravity determination is working in the 5130~K $\le \Teff \le$ 6600~K and 3.12 $\le$ log~g $\le$ 4.66 range. An exception is HD\,64090, for which we determined log~g$_{Sp}$ = 4.70 that is 0.13~dex higher than log~g$_{Hip}$. We, thus, do not confirm a temperature trend of $\Delta$log~g(Sp -- Hip) for cool ($\Teff <$ 5300~K) dwarf (log~g $> 4.2$) stars obtained by \citet{Bensby2014}.

 Final slopes of the [Fe/H] vs. $EW$ plots amount to $-0.0002$\,dex/m\AA,  on average. 
Using the derived microturbulence velocities $\vt$ and basic atmosheric parameters $\Teff$, log~g, and [Fe/H], we built up the approximation formula:

\begin{equation}
 \vt = -0.21 + 0.06 \times {\rm [Fe/H]} + 5.6 \times (\Teff/10^4) - 0.43 \times {\rm log~g}.\label{formula2}
\end{equation}

 Microturbulence velocity grows towards higher \teff\ and lower surface gravity. The metallicity dependence is weak. Similar relations were previously
found by other authors, for example \citet{Nissen1981, Edvardsson1993A&A...275..101E, ap2004}, and \citet{Ramirez2013}.  
In the first three studies theirs formulae did not include the [Fe/H] term because these studies dealt with the limited metallicity range.
Using the formula (\ref{formula2}) results in $\vt$ = 1.1\,\kms\ and 1.4\,\kms\ for the solar atmospheric parameters and 6000/4.0/$-1.0$, respectively. For comparison, the corresponding values are 1.1\,\kms\ and 1.5\,\kms\ from the \citet{Ramirez2013} formula.

\subsection{Notes on individual stars}\label{Sect:notes}

We provide additional comments on stars that turned out to be interesting for one or the other reason in the course of the analysis.

{\it HD\,59984 and HD\,105755}, with [Fe/H] = $-0.69$ and $-0.73$, respectively, may be the most metal-poor thin-disk stars. Both kinematics, $v_{pec}$ = 25\,\kms\ and 28\,\kms, and an age of 8~Gyr for both stars support their thin-disk status.

{\it HD\,106516}:  We failed to obtain self-consistent stellar parameters for this star. On the one hand, the star exhibits a typical thick-disk kinematics, $v_{pec}$ = 108\,\kms, and chemical signatures, [Mg/Fe] = 0.38 and [Fe/H] = $-0.73$.
On the other hand, its age was estimated to be 6~Gyr and unlikely older than 9~Gyr, identifying the star as a thin-disk star.
 According to \citet{2001AJ....122.3419C}, it is a single-lined spectroscopic binary, with the period $P$ = 843.9 days. 

{\it HD\,134169}: kinematics, $v_{pec}$ = 5\,\kms\ identifies it as a thin-disk star.
However, HD\,134169 exhibits a typical thick-disk chemical signatures, [Mg/Fe] = 0.32 and [Fe/H] = $-0.78$, and old age, $\tau \simeq$ 11~Gyr. Enhancement of Al relative to Fe, with [Al/Fe] = 0.54, found by \citet{2006A&A...451.1065G} also suggests a thick-disk origin. 

{\it BD\,+37$^\circ$ 1458}: this is a halo star. With spectroscopic \teff\ = 5500~K, log~g = 3.70, and [Fe/H] = $-1.95$, the star sits on the evolutionary track of $M = 0.67 M_\odot$ at $\tau$ = 24~Gyr. The younger age can only be obtained for higher \teff\ and log~g. However, there is no ground for a substantial revision of atmospheric parameters. The star's temperature is well fixed by several studies. \citet{Alonso1996irfm,2005ApJ...626..446R}, and \citet{GH2009A&A...497..497G} derived $T_{\rm IRFM}$ = 5510~K, 5516~K, and 5582~K, respectively. \citet{1994A&A...291..895A} determined $\Teff$ = 5450~K from the Balmer line wing fits. With $\Teff$ = 5500~K, log~g$_{Hip}$ = $3.41\pm0.18$ is even lower than the spectroscopically derived value. 
For this star we choose the spectroscopic stellar parameters as the final ones.

\subsection{Uncertainties in derived atmospheric parameters}\label{Sect:uncertain}

The following approaches were applied to evaluate the uncertainties in the derived atmospheric parameters. 

For the benchmark stars we adopted the $T_{\rm IRFM}$ errors as indicated by the original sources. 
Statistical error of the Hipparcos parallax based surface gravity was computed as the quadratic sum of errors of the star's parallax, effective temperature, mass, visual magnitude, and bolometric correction: 
$$\sigma_{\rm log g}^2 = (2 \sigma_{\log\pi})^2 + (4 \sigma_{\log T})^2 + \sigma_{\log M}^2 + (0.4 \sigma_V)^2 + (0.4 \sigma_{\rm BC})^2 .$$ 

\noindent The uncertainties in $V$ magnitude and bolometric correction together contribute less than 0.01~dex to the total log~g error. 
 For \teff\ the contribution nowhere exceeds 0.03~dex. Stellar masses were well constrained using the evolutionary tracks, with an uncertainty of no more than 0.1~$M_\odot$, resulting in 0.04~dex contribution to $\sigma_{\rm log g}$. 

\begin{figure}
\epsscale{1.0}
\plotone{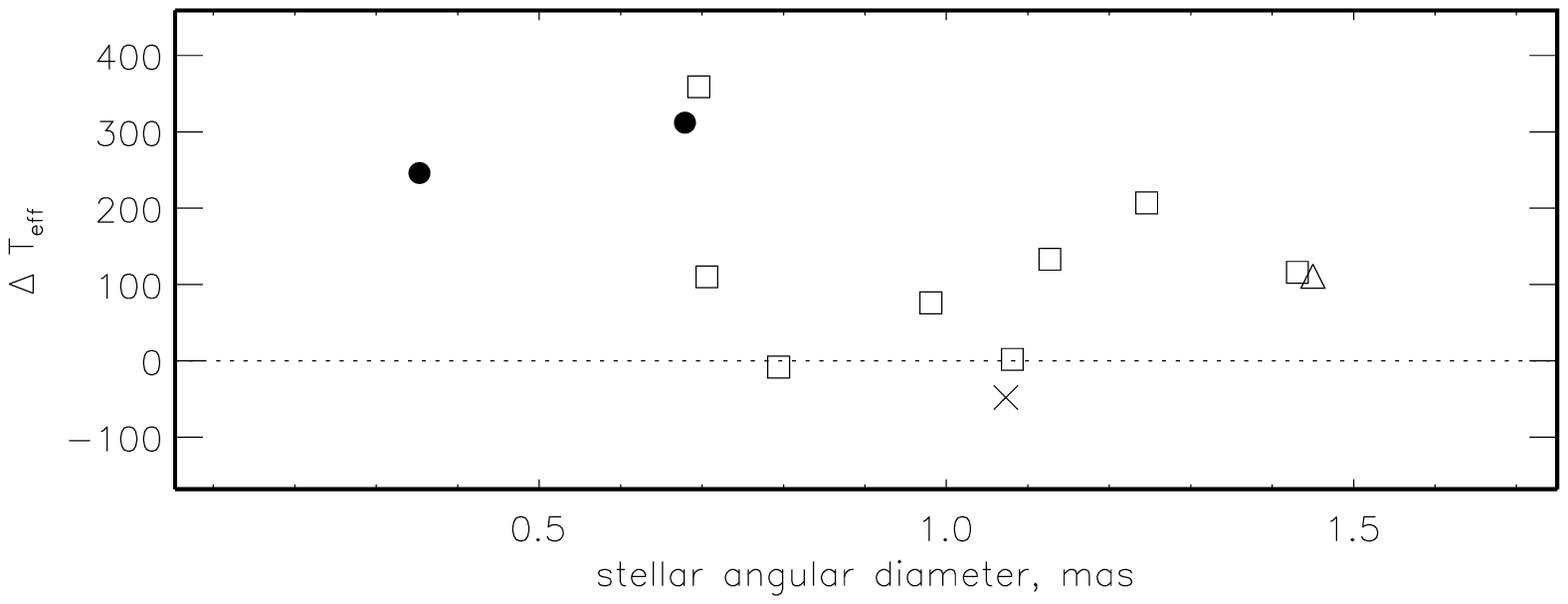}
\plotone{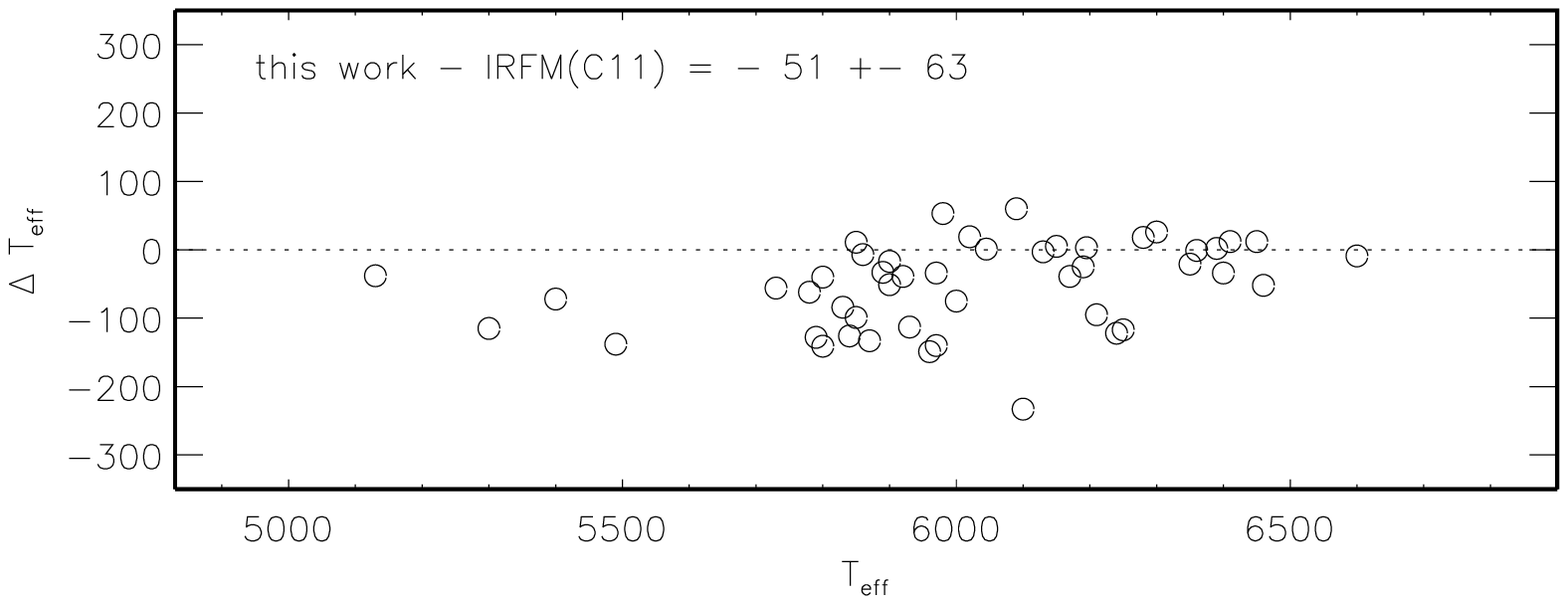}
\plotone{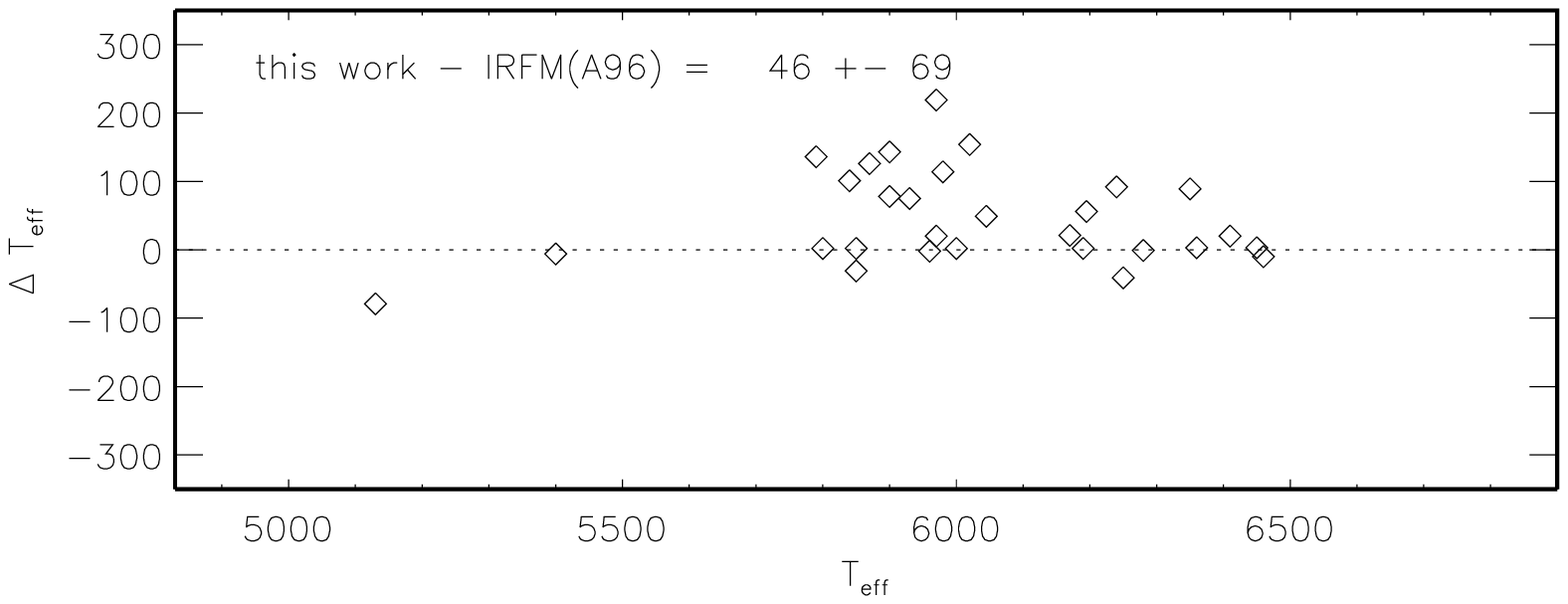}
\caption{Differences in effective temperature between this study and the literature data.  The top panel corresponds to interferometric measurements of \citet[][squares]{Boyajian2012,Boyajian2013}, \citet[][filled circles]{Creevey2012,Creevey2014}, \citet[][triangle]{North2009},  and \citet[][cross]{vonBraun2014}. See sect.\,\ref{Sect:comparisons} for more details. The middle and bottom panels use the IRFM temperatures from \citet{Casagrande2011} and \citet{Alonso1996}, respectively.  
\label{this_irfm}}
\end{figure}

In Fig.\,\ref{this_irfm} our final $\Teff$s are compared with the IRFM temperatures from C11 and A96. For 47 stars in common with C11, our values are  51~K, on average, lower.
In contrast,  this study determined 46~K higher temperatures compared with that of A96 for 29 common stars. 
We note very similar statistical errors of 63~K and 69~K for the temperature differences
 (this work -- C11) and (this work -- A96), respectively. Based on these comparisons, we estimate the systematic and statistical errors of $\Teff$ for the 31 stars with the spectroscopically derived atmospheric parameters to be 50~K and 70~K.

For 20 stars with the spectroscopic surface gravities, log~g$_{Sp}$, their {\sc Hipparcos} parallaxes were measured with a relative parallax error less than 10\,\%, and we calculated reliable log~g$_{Hip}$ values. As shown in Fig.\,\ref{sp_hipp}, log~g$_{Sp}$ -- log~g$_{Hip}$ = $0.008\pm0.037$~dex. This led us to estimate the uncertainty in log~g$_{Sp}$ to be 0.04~dex.

Statistical error of [Fe/H] was defined by the dispersion, $\sigma$, for lines of Fe~II in given star. 

Statistical error of the microturbulence velocity was adopted to be common for the whole stellar sample, and it was defined by the dispersion in the single $\vt$ measurements about the relation (\ref{formula2}). It amounts to 0.14~\kms.

\begin{figure}
\epsscale{1.0}
\plotone{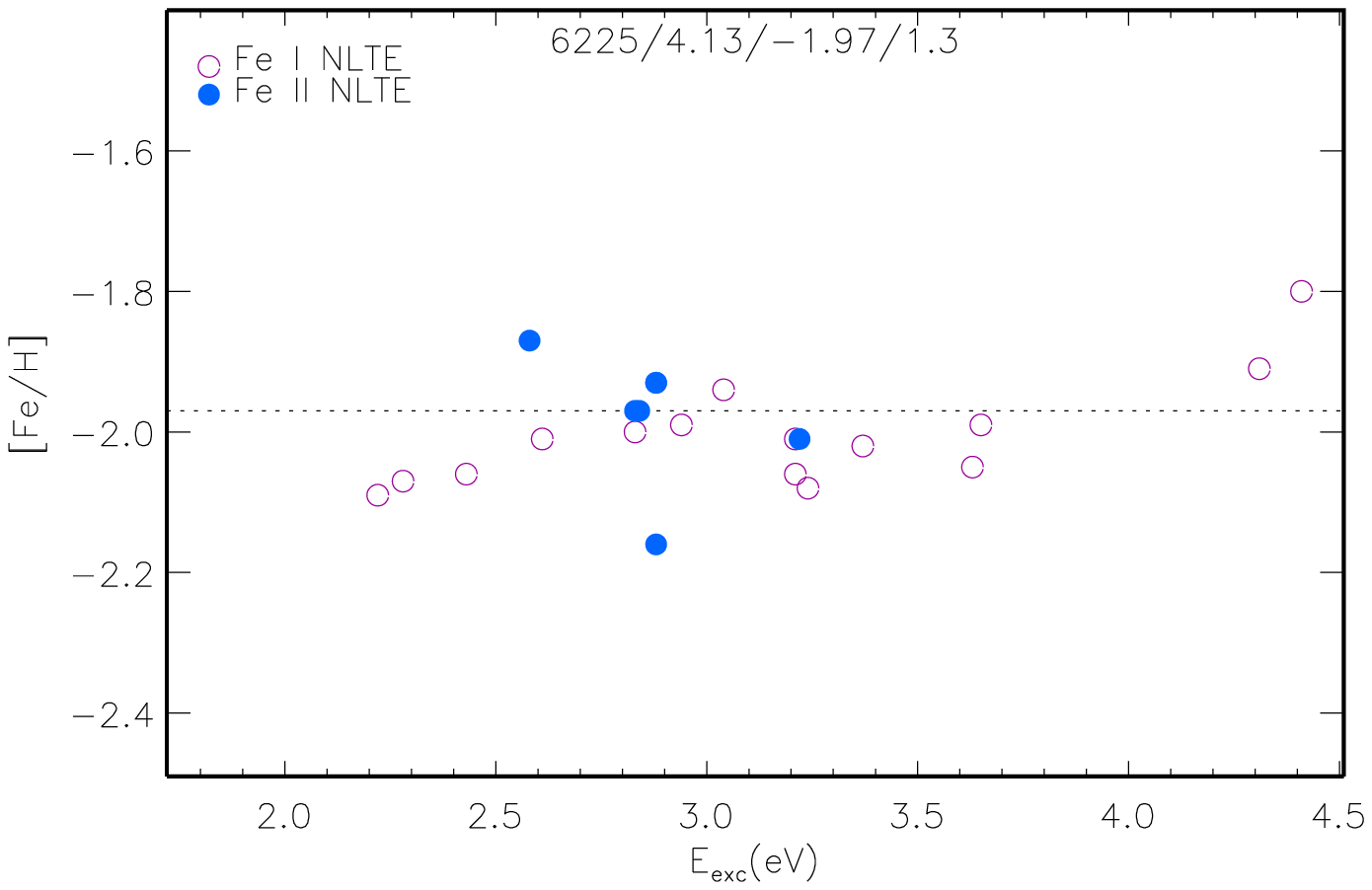}
\plotone{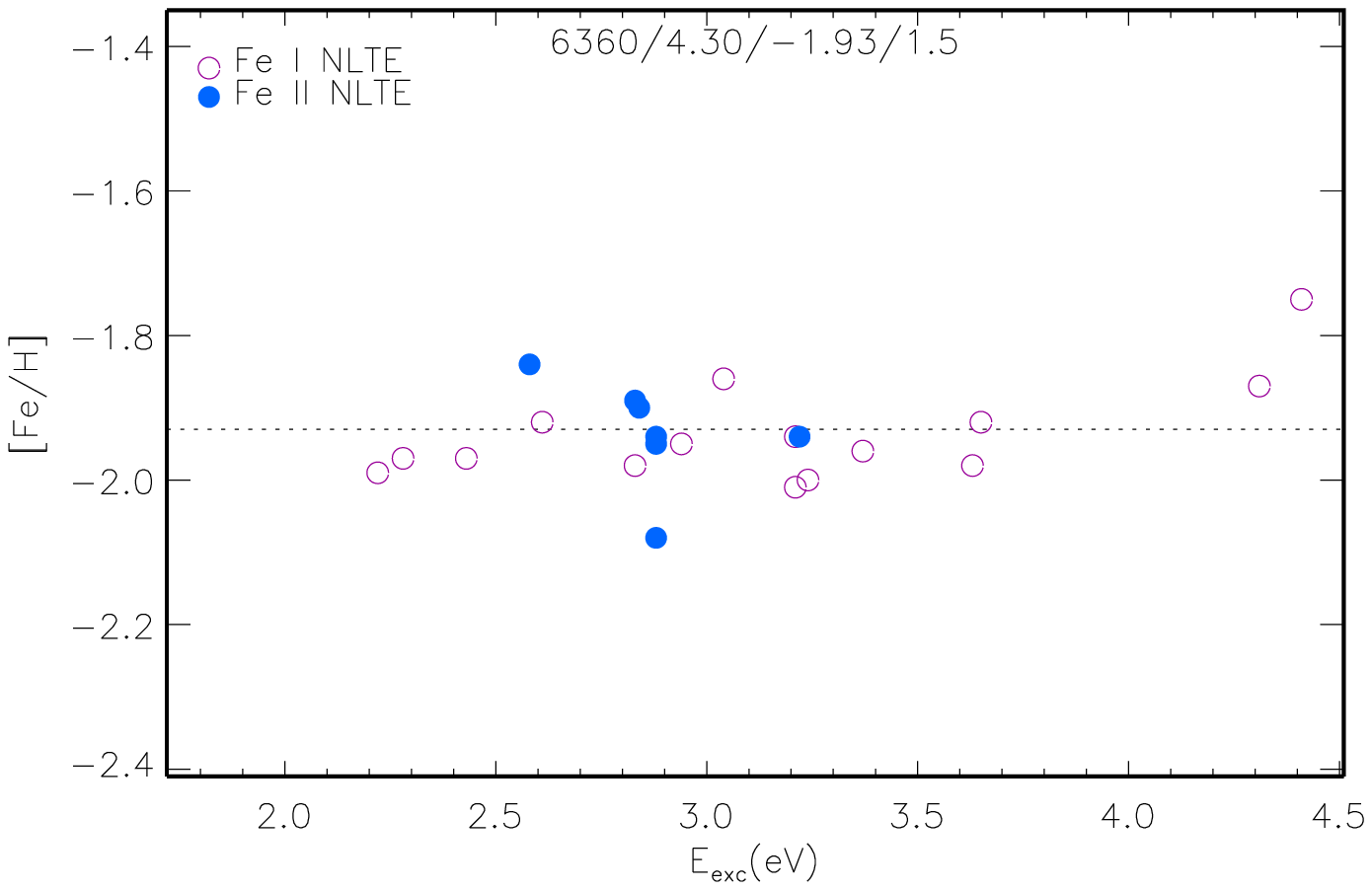}
\caption{Impact of changes in \teff\ and log~g on the NLTE differential abundances derived from the Fe~I (open circles) and Fe~II (filled circles) lines in HD~74000. The top panel corresponds to our final parameters, \teff\ = 6225~K, log~g = 4.13, [Fe/H] = $-1.97$, and the bottom one to \teff\ = 6360~K, log~g = 4.30, [Fe/H] = $-1.93$. The differences in NLTE abundances Fe~I -- Fe~II = $-0.03$~dex and +0.01~dex, respectively. \label{hd74000}}
\end{figure}

It is worth noting that the uncertainty in spectroscopically derived atmospheric parameters can be larger than that quoted above. For example, for the halo star HD\,74000 we found the two sets of stellar parameters that reproduce well the Fe~I/Fe~II ionization and Fe~I excitaion equilibrium (Fig.\,\ref{hd74000}) and fit the appropriate evolutionary tracks. These are $\Teff$ = 6225~K, log~g = 4.13, and [Fe/H] = $-1.97$ suggesting the star's age of 15~Gyr and $\Teff$ = 6360~K, log~g = 4.30, and [Fe/H] = $-1.93$ resulting in $\tau$ = 13.5~Gyr. The first set relied, in fact, on $T_{\rm IRFM}$(A96) and the second one on $T_{\rm IRFM}$(C11). A shift of +135~K in \teff\ leads to +0.17~dex shift in log~g, and both sets of parameters provide an acceptable star's age. No of two sets can be preferred from a comparison of log~g$_{Sp}$ with the corresponding {\sc Hipparcos} parallax based surface gravity, log~g$_{Hip} = 4.16\pm0.16$ for $\Teff$ = 6225~K and log~g$_{Hip} = 4.20\pm0.16$ for $\Teff$ = 6360~K. From analysis of H$_\alpha$ and H$_\beta$ in HD\,74000 \citet{Mashonkina2003} inferred $\Teff$ = 6225~K.
 Two pairs of $\Teff$ / log~g, which provide the Fe~I/Fe~II ionization equilibrium and reasonable star's mass and age, were also obtained for 
HD\,108177 (6100/4.22/$-1.67$ and 6250/4.40/$-1.62$), and G090-003 (6100/3.90/$-2.04$ and 5930/3.80/$-2.10$).
We, therefore, recommend to apply various spectroscopic and non-spectroscopic methods to given star to find an unique solution for the star's \teff~/~log~g.

\section{Comparison with other studies}\label{Sect:comparisons}

For ten stars of our sample, their $\Teff$s were determined in the literature based on measurements of the angular diameters, trigonometric parallaxes, and bolometric fluxes. \citet[][B13]{Boyajian2013} published the most numerous list of the interferometric temperatures, $T_{int}$, based on angular diameters measured with the CHARA array (330~m maximum baseline). 
Figure\,\ref{this_irfm} displays the temperature differences between this study and interferometric measurements of B13, \citet{North2009,Creevey2012,Creevey2014}, and \citet{vonBraun2014}.  Our values are, on average, higher, with $\Delta\Teff$ = 135 $\pm$ 126~K for ten stars and 78 $\pm$ 81~K, when excluding the two outliers, HD~103095 and HD~140283.  We selected three stars for a detailed comparison. 

{\it HD~102870.} Two successive determinations of \citet{Boyajian2012} and B13 resulted in 
$T_{int}$ = 6132~K and 6054~K, with small temperature errors of 36~K and 13~K, respectively. A downward revision by 78~K was due to employing the different bolometric fluxes. The most recent temperature of B13 is in line with the earlier data of \citet{North2009}, $T_{int}$ = 6059 $\pm$ 49~K.
Surface gravity based on the asteroseismic measurements amounts to log~g$_{seis}$ = 4.11 $\pm$ 0.02 \citep{creevey2013}. With \teff\ = 6060~K and log~g = 4.11, we failed to achieve the ionization equilibrium between Fe~I and Fe~II and obtained Fe~I -- Fe~II = $-0.12$~dex in NLTE. It is worth noting, log~g$_{Hip}$ = 4.14 adopted in this study agrees well with log~g$_{seis}$. For solar-type stars the asteroseismology method depends only weakly on \teff. Indeed, a 200~K temperature difference produces a difference of 0.007~dex in log~g. To obtain consistent NLTE abundances from lines of Fe~I and Fe~II using log~g = 4.14, one needs to have 110~K higher temperature of HD~102870 (Table\,\ref{params}) compared with its $T_{int}$.

{\it HD~103095.} Using the CHARA array measurements, B13 and \citet{Creevey2012} derived $T_{int}$ = 4771 $\pm$ 18~K and 4818 $\pm$ 54~K, respectively. 
We checked \teff\ = 4820~K and log~g$_{Hip}$ = 4.60 with various spectroscopic temperature and gravity indicators. 

(i) The Fe~I/Fe~II ionization equilibrium is not fulfilled, and the NLTE abundance difference amounts to Fe~I -- Fe~II = $-0.27$~dex.

(ii) We compared the NLTE abundances from C~I 9094 \AA, 9111 \AA\ with the carbon abundance from a number of molecular CH bands,
which are known to be sensitive to \teff\ variation. The NLTE method for C~I and atomic and molecular data were taken from \citet{Alexeeva_carbon}. With the 4820/4.60/$-1.3$ model, the abundance difference, C~I -- CH = 0.56~dex, is very large, while it amounts to $-0.06$~dex for our final parameters, \teff\ = 5130~K and log~g = 4.66. 

\begin{figure}
\epsscale{1.0}
\plotone{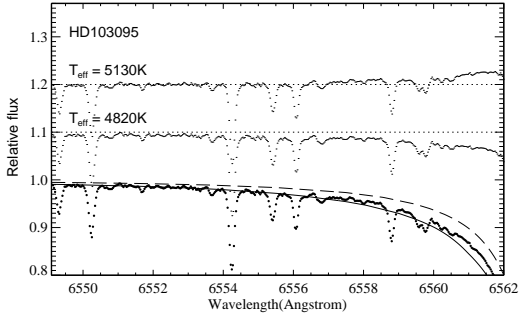}
\caption{Theoretical NLTE flux profiles of H$_\alpha$ calculated with \teff\ = 4820~K (dashed curve) and \teff\ = 5130~K (continuous curve) compared to the observed FOCES spectrum of HD\,103095 (bold dots). The differences between observed and calculated spectra, (O - C), are shown in the upper part of the panel.
\label{Halpha_HD103095}}
\end{figure}

(iii) Effective temperature can also be constrained from analysis of the H$_\alpha$ line wings 
(Fig.\,\ref{Halpha_HD103095}). The theoretical NLTE profiles of H$_\alpha$ were computed following \citet{Mashonkina2008}. 
It can be seen that \teff\ = 4820~K leads to shallower wings of H$_\alpha$ compared with the observations, but \teff\ = 5130~K to deeper core-to-wing transition region. The best fit was achieved for \teff\ = 5030~K. It is worth noting, \citet{Cayrel2011} obtained 100~K, on average, lower temperatures from the H$_\alpha$ line wings compared with accurate effective temperatures from the apparent angular diameter for the eleven FGK-type stars in the $-0.7 \le$ [Fe/H] $\le 0.2$ range. In contrast, $T_{int}$ = 4771~K and 4818~K of HD\,103095, as determined by B13 and \citet{Creevey2012}, respectively, are more than 200~K lower than the temperature from H$_\alpha$. 

Thus, no spectroscopic temperature indicator supports the literature data on $T_{int}$ for HD\,103095.

\begin{figure}
\epsscale{1.0}
\plotone{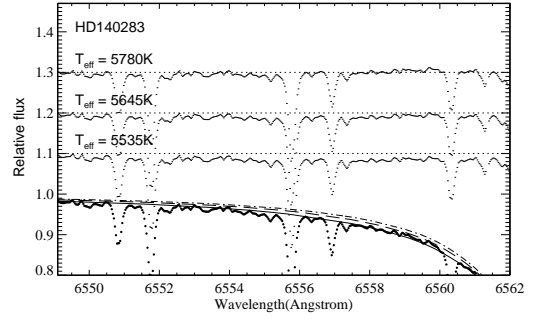}
\caption{Theoretical NLTE flux profiles of H$_\alpha$ calculated with \teff\ = 5535~K (dash-dotted curve), 5645~K (dashed curve), and 5780~K (continuous curve) compared to the observed VLT/UVES spectrum of HD\,140283 (bold dots). The (O - C) values are shown in the upper part of the panel.
\label{Halpha_HD140283}}
\end{figure}

{\it HD\,140283.} \citet{Creevey2014} derived $T_{int}$ = 5534 $\pm$ 103~K and 5647 $\pm$ 105~K assuming zero-reddening and A$_V$ = 0.1$^m$, respectively. It is worth noting, HD\,140283 is a nearby star, at a distance of 58~pc from the Sun, and an interstellar absorption of A$_V$ = 0.1$^m$ is unlikely produced.  Of the two temperatures $T_{int}$ = 5534~K should be preferred, nevertheless we checked both using the same spectroscopic indicators as for HD\,103095. The NLTE abundance differences Fe~I -- Fe~II were found to be $-0.18$~dex and $-0.09$~dex, respectively. For both temperatures an abundance difference between carbon atomic and molecular lines is large, with (C~I -- CH) = 0.69~dex and 0.32~dex, while consistent abundances, (C~I -- CH) = 0.00~dex, were obtained for our final parameters, \teff\ = 5780~K and log~g = 3.70.
Figure\,\ref{Halpha_HD140283} shows theoretical profiles of H$_\alpha$ in the three different model atmospheres compared with the observed spectrum of HD\,140283. It is evident that $T_{int}$ = 5534~K is too low and it does not fit any spectroscopic indicator of \teff. 

 It is worth noting, both HD~103095 and HD~140283 have rather small angular diameters, $\theta_{int}$ = 0.679 $\pm$ 0.007~mas \citep{Creevey2012} and 0.353 $\pm$ 0.013~mas \citep{Creevey2014}, respectively, which can be overestimated resulting in underestimated temperatures. Indeed, \citet{2014MNRAS.439.2060C} suspected systematic trends in the \citet[][B12]{Boyajian2012} dataset that was also based on the CHARA array interferometric measurements. They obtained the differences ($T_{\rm IRFM}$(C11) - $T_{int}$(B12)) growing towards smaller angular diameter and reaching +200~K, on average, at $\theta_{int}$ = 0.8~mas. 

\begin{figure}
\epsscale{1.0}
\plotone{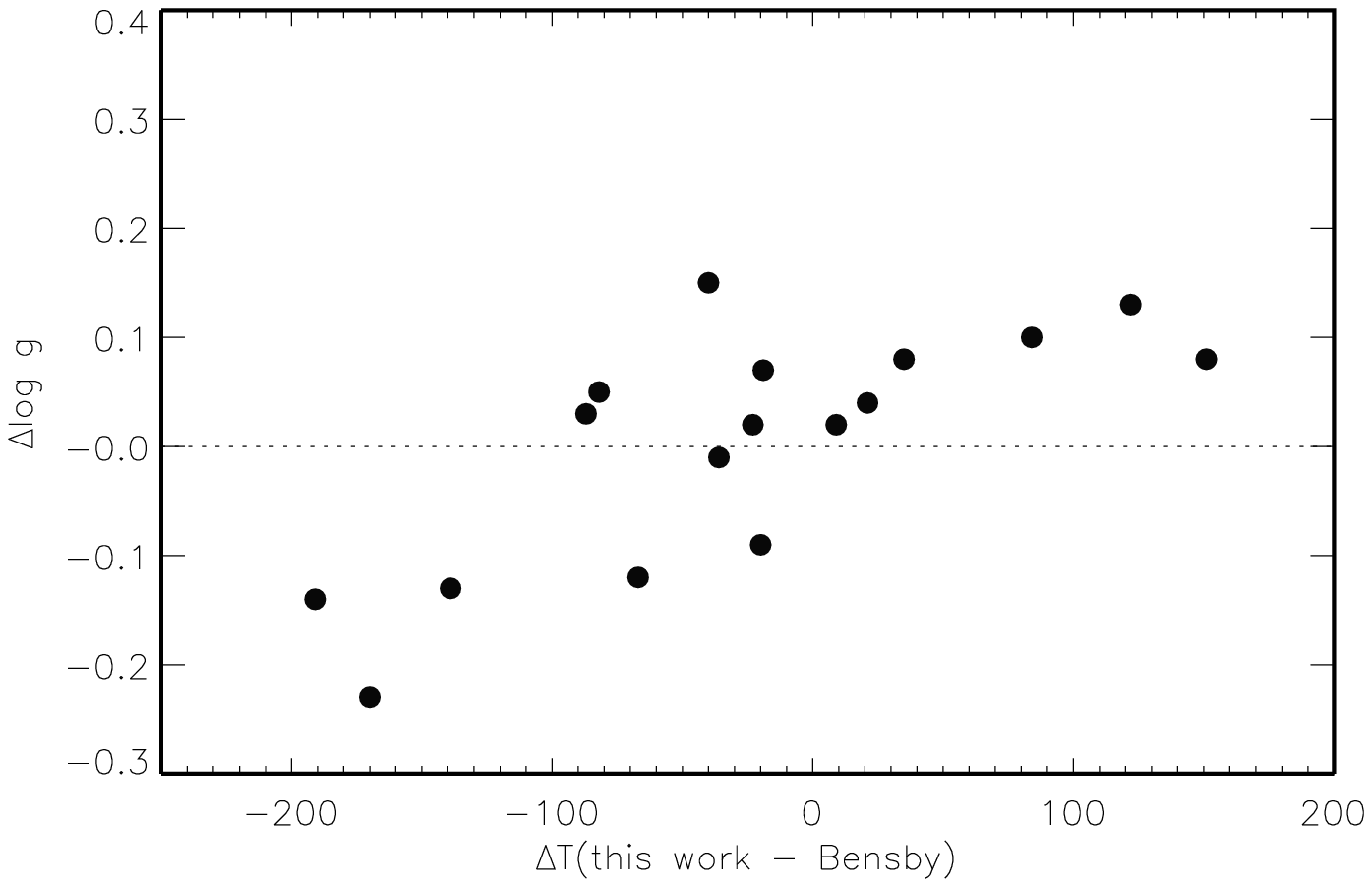}
\plotone{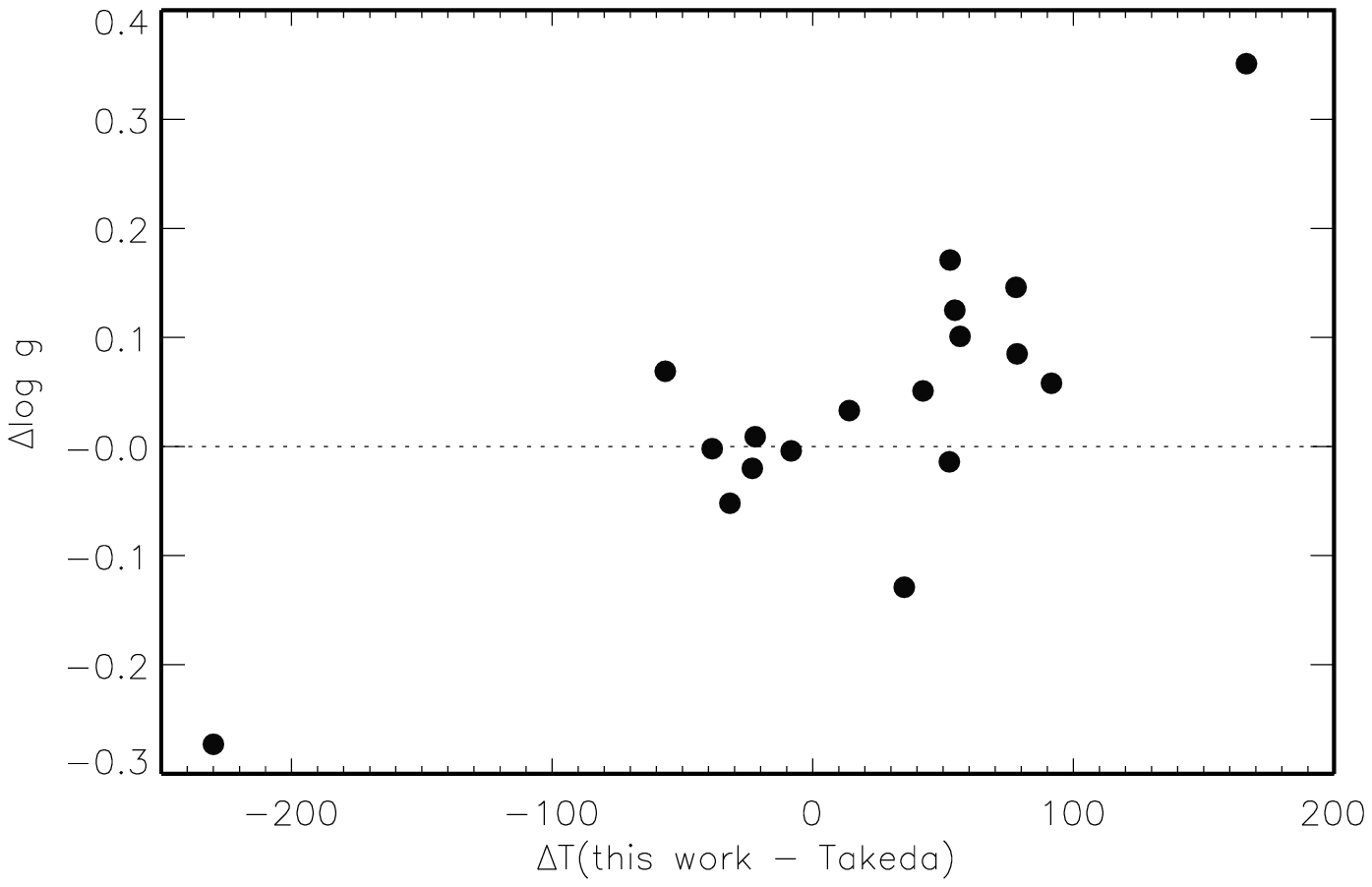}
\plotone{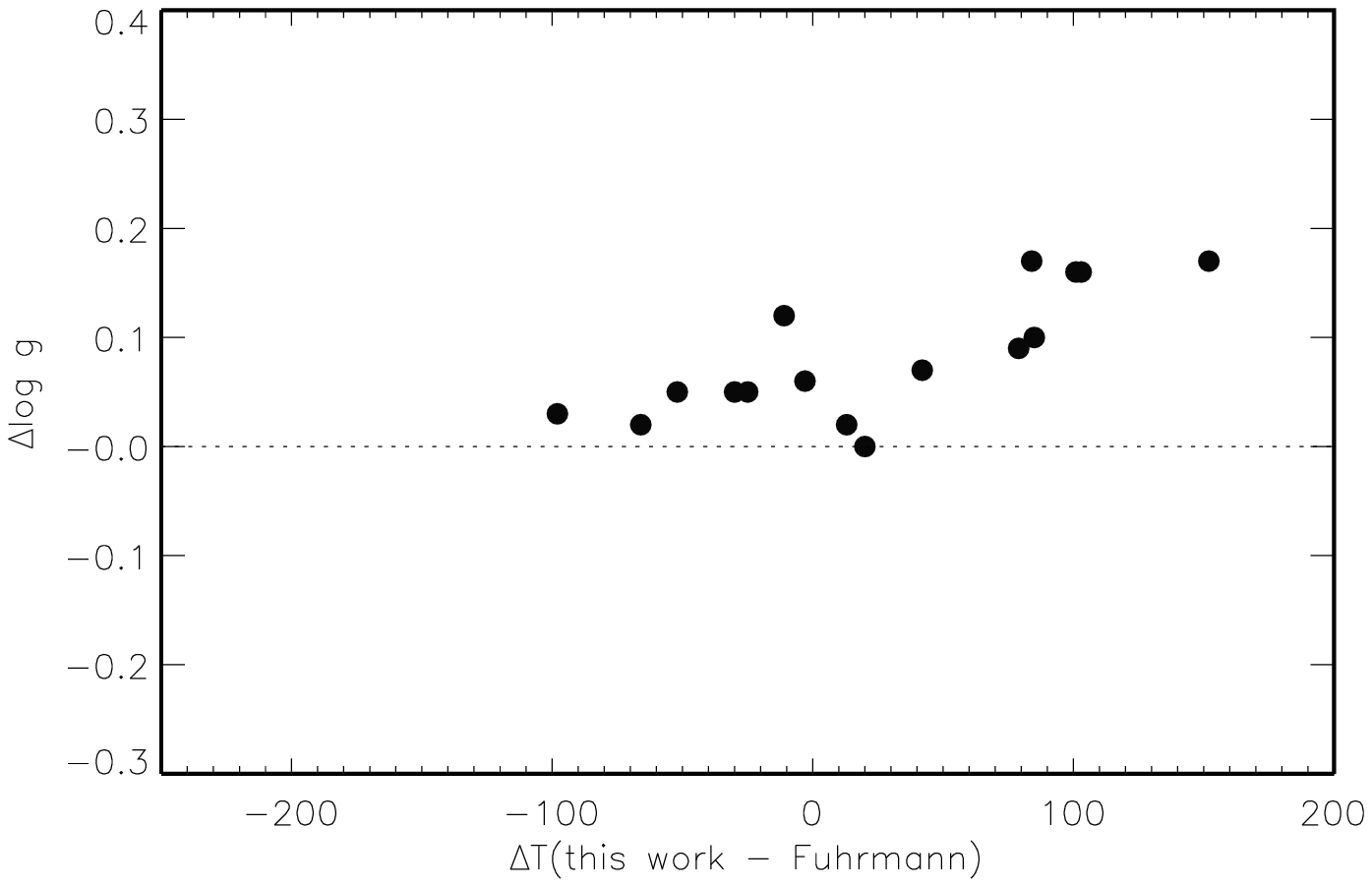}
\caption{Differences in effective temperature and surface gravity between this study and the spectroscopic determinations by \citet{Bensby2014} (top panel), \citet{Takeda2005} (middle panel), and \citet{Fuhrmann1998,Fuhrmann2004} (bottom panel).
\label{this_spectr}}
\end{figure}

 Asteroseismic measurements of surface gravity were made for one more star of our sample, HD~49933 \citep{Kallinger2010}. They resulted in log~g$_{seis}$ = 4.22, which is only 1.4$\sigma_g$ higher, than our value. 

Spectroscopic stellar parameters were determined in a number of recent studies. We selected three of them for comparison. \citet{Bensby2014} used lines of Fe~I and Fe~II to derive temperatures and gravities of the extended stellar sample. The LTE abundances from individual lines were corrected using the grid of the NLTE abundance corrections from \citet{lind2012}. For nearby stars with very good {\sc Hipparcos} parallaxes, \citet{Bensby2014} calculated also the log~g$_{Hip}$ values and found systematic discrepancies between the surface gravities from the two methods for the stars with log~g $>$ 4.2 and $\Teff <$ 5650~K. Therefore, they applied empirical corrections to the atmospheric parameters from ionization balance. It should be reminded, we obtained no temperature trend of $\Delta$log~g(Sp -- Hip) for our program stars (Sect.\,\ref{Sect:parameters}).
Figure\,\ref{this_spectr} (top panel) shows the differences in $\Teff$ and log~g for 17 stars in common with \citet{Bensby2014}. On average, $\Delta\Teff$ = $-34\pm$87~K, $\Delta$log~g = $-0.02\pm$0.11, and $\Delta$[Fe/H] = $-0.02\pm$0.11 in the sense ``this work minus Bensby''. The two independent research provide consistent stellar parameters, with the small systematic shifts and the statistical errors being typical of stellar parameter determinations. However, for some individual stars the discrepancies are uncomfortably large. For example, $\Delta\Teff$ exceeds 100~K for five stars and $\Delta$log~g $>$ 0.2 for HD\,22879. We note, in particular, the differences of different sign for two well-studied very metal-poor stars HD~84937 and HD\,140283, with $\Delta\Teff = -190$~K and +120~K and $\Delta$log~g $= -0.14$ and +0.13, respectively. This can be due to large uncertainty in evaluating a slope of the log~A(Fe~I) versus $\eexc$ trend, where the total number of investigated lines is limited. Since the Fe~I/Fe~II ionization equilibruim depends on not only surface gravity, but also \teff, an overestimation (underestimation) of \teff\ produces an upward (downward) shift in log~g. 

\citet{Takeda2005} derived stellar \teff\ and log~g from lines of Fe~I and Fe~II under the LTE assumption. However, all the 18 stars in common with our work lie in the $-1.3 \leq$ [Fe/H] $\leq$ 0.2 metallicity range, where the departures from LTE for Fe~I and Fe~II are small. Therefore, the obtained differences in $\Teff$ and log~g (Fig.\,\ref{this_spectr}, middle panel) cannot be caused by applying different line-formation treatments. The data in Fig.\,\ref{this_spectr} suggest, most probably, that the uncertainty in the derived effective temperature was directly translated to the uncertainty in surface gravity.

\citet{Fuhrmann1998,Fuhrmann2004} applied different spectroscopic approaches to derive \teff\ from the Balmer line wing fits and log~g from the  Mg~Ib line profile fits under the LTE assumption. For 16 stars in common with this work, the mean differences in effective temperature and iron abundance, $\Delta\Teff$ = 29 $\pm$ 71~K (Fig.\,\ref{this_spectr}, bottom panel) and $\Delta$[Fe/H] = 0.00 $\pm$ 0.07, do not exceed the error bars. The only star, HD\,30743, was found with $\Delta\Teff >$ 100~K. However, this study obtains higher surface gravities, with $\Delta$log~g = 0.08 $\pm$ 0.07. The discrepancy can be explained, in part, by using the LTE approach in \citet{Fuhrmann1998,Fuhrmann2004}. In the stellar parameter range, with which we concern, Mg~I is subject to overionization resulting in weakened Mg~Ib line. Ignoring the departures from LTE leads to an underestimation of the derived gravity.

\section{Conclusions}\label{Sect:Conclusions}

The sample of 51 nearby FG dwarf stars uniformly distributed over the $-2.60 < {\rm [Fe/H]} < +0.20$ metallicity range was selected for a systematic NLTE investigation of the Galaxy chemical evolution. A membership of individual stars to the particular galactic stellar population, namely the thin disk, the thick disk, and the halo, was mainly identified using the star's kinematics and for few stars using also the chemical signatures, [Fe/H] and [Mg/Fe]. A combination of the photometric and spectroscopic methods was applied to derive a homogeneous set of the stellar atmosphere parameters: \teff, log~g, [Fe/H], and $\vt$. Our spectroscopic analyses took advantage of employing the high-resolution ($R \ge$ 60\,000) observed spectra and NLTE line formation for Fe~I and Fe~II in the classical 1D model atmospheres. Spectroscopic method of \teff/log~g determination was tested using the 20 benchmark stars, for which there are multiple measurements of the IRFM effective temperature and their Hipparcos parallax error is less than 10\,\%. An efficiency of poorly known inelastic Fe+H collisions in the SE of Fe~I-Fe~II was estimated empirically from analysis of their different influence on lines of Fe~I and Fe~II in the MP benchmark stars. We found abundances from the two ionisation stages to be consistent within 0.06~dex for every star, when applying \kH\ = 0.5.
 
Stellar parameters of the remaining 31 stars were obtained spectroscopically from the NLTE analysis of the iron lines.
For lines of Fe~II in our program stars the NLTE abundance corrections do not exceed 0.01~dex in absolute value. The deviations from LTE for Fe~I grow towards higher $\Teff$ and lower [Fe/H] and log~g. For stars with either [Fe/H] $\ge -0.75$, or \teff\ $\le$ 5750~K, or log~g $\ge$ 4.20, the difference in average abundance between NLTE and LTE was found to be less than 0.06~dex, which translates to a shift of less than 0.1~dex in log~g. Since NLTE leads to weakened lines of Fe~I, but it does not affect lines of Fe~II until the extremely low metallicities, the ionization equilibrium Fe~I/Fe~II is achieved for higher gravities in NLTE than in LTE.
The NLTE analysis is crucial for the VMP turn-off and subgiant stars. Indeed, the shift in log~g between NLTE and LTE reaches +0.45~dex for BD$-13^\circ$~3442, with \teff\ = 6400~K, log~g(NLTE) = 3.95, and [Fe/H] = $-2.62$.

The obtained effective temperatures and surface gravities were checked by comparing the star's position in the log~g -- \teff\ plane with the theoretical evolutionary track of given metallicity and $\alpha$-enhancement in the \citet{Yi2004} grid. Most stars reveal self-consistent data on star's mass, age, and kinematics.

Our final effective temperatures lie exactly in between the $T_{\rm IRFM}$ scales of A96 and C11, with a systematic shift of +46~K and $-51$~K, respectively. We estimate the \teff\ statistical error to be about 70~K. Surface gravities obtained from the Fe~I/Fe~II and Hipparcos parallax methods were found to agree well. We do not confirm a temperature trend of $\Delta$log~g(Sp -- Hip) reported by \citet{Bensby2014} for the log~g $>$ 4.2 and $\Teff <$ 5650~K stars. 

We recommend to apply a line-by-line differential approach relative to the Sun,
to minimise the effect of the uncertainty in $gf$-values on the final
results. It is quite efficient for the [Fe/H]$ > -1.5$ stars. However, in the more metal-poor stars, the line-to-line scatter for Fe~I has a similar magnitude for the differential and absolute abundances. This is, probably, due to uncertainties in $C_{6}$-values for the high-excitation lines. We note, in particular, Fe~I 5367\AA\ ($\eexc$ = 4.41~eV) and Fe~I 5383 \AA\ ($\eexc$ = 4.31~eV).

It was found that none of checked spectroscopic temperature indicators, such as Fe~I versus Fe~II, C~I versus CH, and the H$_\alpha$ line wings, do support interferometric effective temperatures of HD~103095, $T_{int}$ = 4818~K $\pm$ 54~K \citep{Creevey2012}, and HD~140283, $T_{int}$ = 5534 $\pm$ 103~K \citep{Creevey2014}. Both stars have rather small angular diameters, and their measurements can be affected by some systematic trends, as suspected by \citet{2014MNRAS.439.2060C} for the \citet{Boyajian2012} dataset.

We conclude that the NLTE analysis of lines of iron in the two ionisation stages, Fe~I and Fe~II, is efficient in deriving reliable atmospheric parameters for the FG-type dwarf stars in the broad metallicity range, down to [Fe/H] = $-2.6$. The stellar parameter determinations would benefit from using the complementary data on interferometric and IRFM temperatures, trigonometric parallaxes, and asteroseismology measurements. We also recommend to combine different spectroscopic temperature and gravity indicators and check the obtained \teff/log~g values with the theoretical evolutionary tracks / isochrones.

The obtained accurate atmospheric parameters will be used in the forthcoming papers to determine NLTE abundances of important astrophysical elements from lithium to europium and improve observational constraints on the chemo-dynamical models of the Galaxy evolution. We also plan to extend our sample toward lower metallicity and surface gravity.

\begin{acknowledgements}

The authors thank  Prof. Debra Fisher for carrying out some of the observations, Oleg Kochukhov and Vadim Tsymbal for providing the codes binmag3 and synthV-NLTE at our disposal and Klaus Fuhrmann for providing the FOCES spectra.  Y.Q. Chen acknowledges the observation of ESPaDOnS/CFHT telescope for two stars obtained through the Telescope Access Program (TAP, 2011B), which is funded by the National Astronomical Observatories.
This study was supported by the Russian Foundation for Basic Research (grant 14-02-91153 and 14-02-31780), the National Natural Science Foundation of China (grants 11390371, 11233004, 11222326, 11103034, 11473033, 11473001), and by the National Basic Research Program of China (grant 2014CB845701/02/03). 
We made use the Simbad, MARCS, and VALD databases.

 We thank the anonymous referee for comments that helped to improve the manuscript.

\end{acknowledgements}

\clearpage

\begin{deluxetable}{rrrllll}
\tabletypesize{\scriptsize}
\centering
\tablecaption{\label{Tab:observations} Characteristics of observed spectra.} 
\tablehead{
 \colhead{HD,BD} & \colhead{V} & \colhead{[Fe/H]} & \colhead{S/N} & \colhead{$t_{exp}$, s} & \colhead{N$_{spec}$} & \colhead{Date (yyyy-mm-dd)}
}
\startdata
       &         &       &  \multicolumn{4}{l}{{\bf Shane/Hamilton}} \\
  19373 &   4.05 &   0.10  &    234, 228   & 30.00, 30.00 & 2 & 2012-01-09, 2012-01-09   \\
  22484 &   4.28 &   0.01  &    171, 160   & 30.00, 30.00 & 2 & 2012-01-09, 2012-01-09   \\ 
  22879 &   6.74 &  -0.84  &    170, 195   & 360.00, 360.00 &  2 & 2012-01-09, 2012-01-09  \\  
  24289 &   9.96 & -1.94   &   112, 128    & 2700.00, 2700.00    & 2 & 2012-01-09, 2012-01-09 \\
  30562 &   5.77 &   0.17  &    180, 190   & 120.00, 120.00 &  2 & 2012-01-09, 2012-01-09  \\
  30743 &   6.26 & -0.44   &   188, 184           &  240.00, 240.00   & 2 & 2012-01-09,  2012-01-09\\
  34411 &   4.70 &   0.01  &    220, 218, 209F$^1$ & 45.00, 45.00, 84.19 & 3 & 2012-01-11, 2012-01-11, 2012-01-07 \\
  43318 &   5.65 & -0.19 &   210, 212, 183F     & 100.00, 100.00,  215.14   & 3 & 2012-01-11, 2012-01-11, 2012-01-07\\
  45067 &   5.90 & -0.16 &   230, 235           & 180.00, 180.00  & 2 & 2012-01-09, 2012-01-09 \\
  45205 &   8.00 & -0.87 &   140, 177           & 1200.00, 1800.00  & 2 & 2012-01-09, 2012-01-09 \\
  49933 &   5.78 &  -0.47  &  216, 203           &  659.71, 135.0     & 2  & 2012-01-08, 2012-01-10\\
  52711 &   5.93 & -0.21 &   256, 163F          & 160.00, 119.18   & 2 & 2012-01-10,2012-01-06  \\
  58855 &   5.36 & -0.29 &   224, 167F          & 150.00, 75.13   & 2 & 2012-01-09, 2012-01-06  \\
  59374 &   8.50 &  -0.88  &    122F, 131F  & 1174.47, 812.31 & 2  & 2012-01-07, 2012-01-06\\
  59984 &   5.93 &  -0.69  &    182F, 165F  &  269.17, 229.17 & 2 & 2012-01-07, 2012-01-06 \\
  62301 &   6.75 & -0.70 &   197, 158F          & 450.00,  238.20   & 2 & 2012-01-09,  2012-01-06 \\
  64090 &   8.30 &  -1.73  &    280, 136F   & 2500.00, 687.35 & 2 & 2012-12-26, 2012-01-06  \\
  69897 &   5.10 &  -0.25  &    186, 180F   &  240.00, 66.18 & 2 & 2012-01-08, 2012-01-06 \\
  74000 &   9.67 & -1.97 &   144, 142, 74F      & 2400.00, 2400.00, 1454.51 & 3 & 2012-01-10, 2012-01-10, 2012-01-06 \\
  76932 &   5.86 & -0.98 &   170, 181F, 210F    & 612.32, 258.23, 489.24  & 3 & 2012-01-08, 2012-01-06, 2012-01-07 \\
  82943 &   6.54 &  0.19 &   214, 170F          & 270.00, 342.25  & 2  & 2012-01-10, 2012-01-06 \\
  84937 &   8.28 &  -2.16 &  122, 153, 95 & 1475.81, 1800.0,  846.41 & 3 & 2012-01-08, 2012-01-09, 2012-01-06 \\
  89744 &   5.74 &  0.13 &   220, 165F          & 135.00, 121.16  & 2 & 2012-01-10, 2012-01-06 \\
  90839 &   4.83 & -0.18 &   190, 169F          & 60.00, 137.20   & 2 & 2012-01-10, 2012-01-06  \\
  92855 &   7.28 & -0.12 &   171, 163F          & 600.00, 309.20  & 2 & 2012-01-10, 2012-01-06  \\
  94028 &   8.23 &  -1.47  &   118, 172    & 1800.00, 1317.69 & 2 & 2011-03-15, 2011-03-15 \\
  99984 &   5.95 & -0.38 &   181F, 183F         & 517.22, 491.24  & 2 & 2012-01-07, 2012-01-07 \\
 100563 &   5.70 &  0.06 &    171F, 173F        & 413.25, 383.21  & 2 & 2012-01-07, 2012-01-07 \\
 102870 &   3.61 &   0.11  &    208F, 211F   & 77.12,  91.13  & 2 & 2012-01-07, 2012-01-07 \\
 103095 &   6.45 &  -1.26  &    188        &  300.00  & 1 & 2012-01-10 \\
 105755 &   8.59 &  -0.73  &    175, 103F    & 2400.0, 1052.44   & 2 & 2012-01-10, 2012-01-06 \\
 106516 &   6.11 & -0.73 &    151F, 155F        & 434.25, 546.25  & 2 & 2012-01-07, 2012-01-07 \\
 108177 &   9.66 & -1.67 &    60, 111           & 3000.00, 3600.00   & 2 & 2011-03-15, 2011-03-15 \\
 110897 &   6.00 & -0.57 &    260, 170F, 172F   & 600.00, 568.30, 508.23  & 3 & 2012-01-08, 2012-01-07, 2012-01-07 \\
 114710 &   4.26 &   0.06  &    184F, 186F   & 112.13, 129.15   & 2 & 2012-01-07, 2012-01-07 \\
 115617 &   4.74 & -0.10  &   177, 170          & 160.00, 160.00  & 2 & 2012-01-08, 2012-01-08 \\
 134088 &   8.00 & -0.80  &   130              & 1800.00   & 1 & 2011-03-15 \\
 134169 &   7.67 &  -0.78  &    214, 212     & 750.00, 750.00   & 2 & 2012-01-10, 2012-01-10  \\
 138776 &   8.72 &  0.24  &   211              &  1800.00  & 1 & 2012-01-11 \\
 142091 &   4.82 &  -0.07  &    277, 275     & 60.00, 60.00   & 2 & 2012-01-11, 2012-01-11 \\
 142373 &   4.62 & -0.54  &   140, 147         &  240.00, 240.00 & 2 & 2012-01-08, 2012-01-08 \\
 $-04^\circ$3208 & 9.99 &  -2.20  &  60, 66          & 1500.00, 2400.00 & 2 & 2012-01-08, 2012-01-08 \\
 $-13^\circ$3442 & 10.37 &  -2.62  &   103, 102, 104, 102  &   2700.00, 2700.00,   & 4 & 2012-01-09, 2012-01-09,   \\
           &       &         &                       &   2700.00, 2700.00    &   &2012-01-09, 2012-01-09\\
 +24$^\circ$1676 & 10.80 &  -2.44  &   86, 90, 92, 92, 57F  &  2400.00, 2400.00, 2400.00,   & 5 & 2012-01-11, 2012-01-11, 2012-01-11,  \\
           &       &         &                        &    2400.00, 1581.60           &   & 2012-01-11, 2012-01-06  \\
 +29$^\circ$2091 & 10.22 &  -1.91  &   82, 121, 115, 120, 83F & 2400.00, 2700.00, 2700.00,   & 5 & 2012-01-08, 2012-01-10, 2012-01-10,  \\
           &       &         &                          &   2700.00, 1287.46           &   & 2012-01-10, 2012-01-06  \\
 +37$^\circ$1458 & 8.92 &  -1.95  &   235, 105F         &  2500.00, 897.38 & 2 & 2012-01-10, 2012-01-06\\
 +66$^\circ$0268 & 9.88 &  -2.06  &    112, 110, 115 & 2400.00, 2400.00, 2400.00 & 3 & 2012-01-10, 2012-01-10, 2012-01-10  \\
 G090-003  & 10.50 &  -2.04  &   106, 120, 122, 66F &  2700.00, 2700.00,  & 4 & 2012-01-11, 2012-01-11,  \\
           &       &         &                      &   2700.00, 1181.44  &   & 2012-01-11, 2012-01-06  \\
       &         &       &  \multicolumn{4}{l}{{\bf CFHT/ESPaDOnS}} \\
 +07$^\circ$4841 & 10.38 &  -1.46  &    152           & 1470   & 2  \\
 +09$^\circ$0352 & 10.17 &  -2.09  &    160            &  2400   & 2 \\
       &         &       &  \multicolumn{4}{l}{{\bf FOCES}} \\
  22879 &   6.74 &  -0.84  & 200 & 900.00, 1800.00, 1800.00  & 3   &   1996-09-02, 1999-01-02, 1996-10-31  \\ 
  34411 &   4.70 &   0.01  & 200 & 1500.00, 1500.00 & 2 & 1999-08-21, 1998-12-27  \\
  84937 &   8.28 &  -2.16  & 200 & 3600.00, 3300.00 & 2 & 1999-03-01, 1999-03-01  \\
  94028 &   8.23 &  -1.47 & 200  & 2400.00, 2400.00 & 2 & 2000-01-17, 2000-01-17 \\
 103095 &   6.45 &  -1.26  & 200 & 900.00   & 1 & 2000-05-19 \\
 142373 &   4.62 &  -0.54  & 200 & 600.00, 600.00 & 2 & 1998-06-09, 1998-06-09 \\
       &         &       &  \multicolumn{4}{l}{{\bf Others}} \\
  49933$^2$ &   4.15 &  -0.47  &  500       &       &  &  \\
 140283$^3$ &   7.21 &  -2.46  &  200          &    &  & \\
 $-04^\circ$3208$^3$ & 9.99 &  -2.20  &  200         &   & & \\
\enddata
\tablecomments{ 
  \ $^1$ F indicates the observations carried out by Debra Fischer, 
  \ $^2$~3.6-m/HARPS,  $^3$  VLT/UVES. \\
}
\end{deluxetable}

\clearpage

\begin{deluxetable}{rrrrrrrrc}
\tabletypesize{\scriptsize}
\tablecaption{Stellar kinematics and membership to particular galactic stellar population.\label{kin}}
\tablewidth{0pt}
\tablehead{
\colhead{HD, BD} & \colhead{X-8000} & \colhead{Y} & \colhead{Z} & \colhead{U} &
\colhead{V} & \colhead{W} & \colhead{[Mg/Fe]} & \colhead{Stellar} \\
\colhead{ } & \colhead{(pc)} & \colhead{(pc)} & \colhead{(pc)} & \colhead{(\kms)} &
\colhead{(\kms)} & \colhead{(\kms)} & \colhead{NLTE} & \colhead{population} 
}
\startdata
   19373& 10&   -10&   -1&     -75.2&     -16.1&      21.5&   0.00 \scriptsize{$\pm$} 0.03 &      Thin disk \\
   22484& 10&    0&   -9&       1.7&     -15.5&     -42.3&  -0.05 \scriptsize{$\pm$} 0.02 &      Thin disk \\
   22879& 20&    0&  -17&    -109.9&     -88.2&     -41.3&   0.30 \scriptsize{$\pm$} 0.02 &    Thick disk \\
   24289&160&    30& -139&    -116.7&    -172.6&     173.9&   0.21 \scriptsize{$\pm$} 0.04 &     Halo \\
   30562& 20&    10&  -13&     -54.9&     -72.0&     -21.6&   0.04 \scriptsize{$\pm$} 0.05 &     Thin disk \\
   30743& 20&    20&  -18&      24.8&      -5.2&     -22.5&   0.08 \scriptsize{$\pm$} 0.06 &     Thin disk \\
   34411& 10&   0&    0&     -74.5&     -35.4&       4.4&   -0.03 \scriptsize{$\pm$} 0.03 &     Thin disk  \\
   43318& 30&    20&   -5&      47.0&       1.2&     -37.0&  -0.05 \scriptsize{$\pm$} 0.05 &      Thin disk \\
   45067& 30&    20&   -3&     -16.5&     -64.4&      13.7&   0.01 \scriptsize{$\pm$} 0.04 &     Thin disk \\
   45205& 60&   -30&   26&     -35.9&     -89.5&      17.0&   0.39 \scriptsize{$\pm$} 0.03 &    Thick disk \\
   49933& 30&    20&    0&      25.7&     -13.1&      -9.2&   0.16 \scriptsize{$\pm$} 0.10 &     Thin disk \\
   52711& 20&    0&    5&     -18.6&     -77.8&      -9.0&   0.01 \scriptsize{$\pm$} 0.05 &     Thin disk \\
   58855& 20&  0&    9&      25.3&     -15.4&      -3.9&   0.07 \scriptsize{$\pm$} 0.03 &     Thin disk \\
   59374& 40&    20&   14&     -48.5&    -120.3&      -9.8&   0.30 \scriptsize{$\pm$} 0.02 &    Thick disk \\
   59984& 20&    20&    3&       8.0&     -11.4&     -21.2&   0.10 \scriptsize{$\pm$} 0.02 &     Thin disk \\
   62301& 30&    0&   16&      -7.4&    -110.0&     -22.1&   0.26 \scriptsize{$\pm$} 0.03 &    Thick disk \\
   64090& 30&    0&   13&     266.9&    -227.6&     -89.6&   0.24 \scriptsize{$\pm$} 0.04 &     Halo \\
   69897& 20&    0&    9&     -23.9&     -38.7&       7.1&  -0.01 \scriptsize{$\pm$} 0.06 &      Thin disk \\
   74000& 60&    110&   34&     216.8&    -348.0&      62.2&   0.37 \scriptsize{$\pm$} 0.03 &     Halo \\
   76932& 10&    20&    7&     -49.1&     -91.3&      70.0&   0.32 \scriptsize{$\pm$} 0.05 &    Thick disk \\
   82943& 10&    20&   13&      13.3&     -13.5&     -12.6&  -0.05 \scriptsize{$\pm$} 0.05 &      Thin disk \\
   84937& 40&    30&   52&     205.0&    -214.7&      -7.9&   0.21 \scriptsize{$\pm$} 0.08 &     Halo \\
   89744& 20&   0&   33&     -10.6&     -29.9&     -14.3&  -0.02 \scriptsize{$\pm$} 0.04 &      Thin disk \\
   90839& 10&   0&   10&     -13.9&      -1.8&       2.4&   0.02 \scriptsize{$\pm$} 0.03 &     Thin disk \\
   92855& 20&   0&   32&     -42.1&     -22.4&     -13.9&  -0.02 \scriptsize{$\pm$} 0.04 &      Thin disk \\
   94028& 20&    10&   42&     -33.7&    -129.3&      13.5&   0.35 \scriptsize{$\pm$} 0.05 &    Thick disk \\
   99984& 20&   -10&   49&      -6.4&      11.0&     -36.1&   0.02 \scriptsize{$\pm$} 0.03 &     Thin disk \\
  100563& 0&    10&   24&     -14.5&     -20.9&      -9.8&  -0.03 \scriptsize{$\pm$} 0.03 &      Thin disk \\
  102870& 0&    10&   10&      40.4&       3.2&       7.3&  -0.02 \scriptsize{$\pm$} 0.03 &      Thin disk \\
  103095& 0&   0&    9&     278.3&    -157.2&     -14.8&   0.24 \scriptsize{$\pm$} 0.06 &     Halo \\
  105755& 30&   -30&   78&      26.7&       5.3&       8.3&   0.34 \scriptsize{$\pm$} 0.08 &     Thin disk \\
  106516& 0&    10&   18&      53.4&     -73.4&     -57.7&   0.38 \scriptsize{$\pm$} 0.04 &    Thick disk \\
  108177&    -20&    50&   99&     126.7&    -262.9&      32.0&   0.23 \scriptsize{$\pm$} 0.04 &     Halo \\
  110897& 0&   0&   17&     -41.7&       7.0&      75.3&   0.18 \scriptsize{$\pm$} 0.03 &     Thin disk \\
  114710& 0&   0&    9&     -49.9&      11.5&       8.4&  -0.07 \scriptsize{$\pm$} 0.04 &      Thin disk \\
  115617& 0&    0&    6&     -23.3&     -47.8&     -31.3&  -0.10 \scriptsize{$\pm$} 0.02 &      Thin disk \\
  134088&    -30&    0&   26&     -25.6&     -72.4&     -68.7&   0.36 \scriptsize{$\pm$} 0.03 &    Thick disk \\
  134169&    -40&   0&   44&       2.5&      -3.3&      -1.9&   0.32 \scriptsize{$\pm$} 0.02 &     Thin disk \\
  138776&    -50&   0&   45&       6.4&     -56.4&      -5.0&   0.02 \scriptsize{$\pm$} 0.04 &     Thin disk  \\
  140283&    -50&   0&   32&    -249.0&    -253.3&      41.1& 0.23 \scriptsize{$\pm$} 0.03  &     Halo \\
  142091&    -10&   -20&   24&      33.6&     -40.7&     -18.1&   0.07 \scriptsize{$\pm$} 0.05  &    Thin disk \\
  142373&    0&   -10&   12&     -41.6&      10.8&     -68.3&   0.18 \scriptsize{$\pm$} 0.01 &     Thin disk \\
$-04^\circ$ 3208&    -30&    110&  169&    -123.5&    -309.9&    -120.2&  0.21 \scriptsize{$\pm$} 0.06 &     Halo \\
$-13^\circ$ 3442&        &        &     &          &          &          &  0.23 \scriptsize{$\pm$} 0.01   &   Halo \\
+07$^\circ$ 4841&    -50&   -150& -123&    -224.4&    -283.9&     -55.3&  0.43 \scriptsize{$\pm$} 0.05 &      Halo \\
+09$^\circ$ 0352&    120&   -40& -120&    -156.5&    -176.9&      97.7&  0.45 \scriptsize{$\pm$} 0.05 &      Halo \\
+24$^\circ$ 1676&    360&    90&  126&     166.8&    -483.7&     106.4&  0.21 \scriptsize{$\pm$} 0.06 &      Halo \\
+29$^\circ$ 2091& 40&    20&   85&     157.3&    -345.9&      88.4&  0.34 \scriptsize{$\pm$} 0.04 &      Halo \\
+37$^\circ$ 1458&    150&   -10&   26&    -280.3&    -233.0&     -26.0&  0.38 \scriptsize{$\pm$} 0.05 &      Halo \\
+66$^\circ$ 0268 & 40&   -40&   10&    -181.0&    -426.4&     -73.1&   0.15 \scriptsize{$\pm$} 0.07 &     Halo \\
G090-003 &   820&    90&  332&      10.4&    -824.6&    -192.4&  0.25 \scriptsize{$\pm$} 0.05 &      Halo\\
\enddata
\end{deluxetable}

\clearpage

\begin{deluxetable}{rcrrccr}
\tabletypesize{\scriptsize}
\tablecaption{Line data, iron NLTE and LTE abundances from an analysis of the solar spectrum, and equvalent widths ($EW$) of the solar lines. \label{lines}}
\tablewidth{0pt}
\tablehead{
\colhead{$\lambda$, \AA}  & \colhead{$\eexc$} &  \colhead{log $gf$} & \colhead{log $C_6$}& \colhead{log $A$}& \colhead{log $A$} & \colhead{$EW$} \\
 \colhead{ } & \colhead{(eV)} &  \colhead{ } & \colhead{ }& \colhead{NLTE}& \colhead{LTE} & \colhead{m\AA} 
}
\startdata
 Fe I    &      &      &        &        & & \\
 4920.50 & 2.83 &   0.07 & -30.51 & -4.61 & -4.61 & 466.0\\     
 5198.72 & 2.22 &  -2.14 & -31.32 & -4.52 & -4.53 & 102.0\\
 5217.40 & 3.21 &  -1.07 & -30.37 & -4.57 & -4.57 & 130.3\\ 
 5232.94 & 2.94 &  -0.06 & -30.54 & -4.68 & -4.68 & 371.1\\
 5236.20 & 4.19 &  -1.50 & -31.32 & -4.68 & -4.69 &  33.7\\ 
 5242.50 & 3.63 &  -0.97 & -31.56 & -4.47 & -4.47 &  93.4\\
 5281.79 & 3.04 &  -0.83 & -30.53 & -4.68 & -4.69 & 163.6\\
 5324.18 & 3.21 &  -0.10 & -30.42 & -4.54 & -4.54 & 332.5\\
 5367.47 & 4.41 &   0.44 & -30.20 & -4.78 & -4.78 & 170.1\\
 5379.58 & 3.69 &  -1.51 & -31.56 & -4.46 & -4.47 &  64.1\\
 5383.37 & 4.31 &   0.64 & -30.37 & -4.72 & -4.72 & 220.5\\
 5393.17 & 3.24 &  -0.72 & -30.42 & -4.60 & -4.60 & 171.2\\
 5491.83 & 4.19 &  -2.19 & -31.33 & -4.53 & -4.54 &  14.5\\
 5586.76 & 3.37 &  -0.10 & -30.38 & -4.58 & -4.58 & 289.3\\
 5662.52 & 4.18 &  -0.57 & -30.52 & -4.41 & -4.42 & 105.7\\
 5696.09 & 4.55 &  -1.72 & -30.21 & -4.68 & -4.69 &  14.6\\ 
 5705.46 & 4.30 &  -1.36 & -30.47 & -4.61 & -4.62 &  41.2\\ 
 5778.45 & 2.59 &  -3.44 & -31.37 & -4.60 & -4.61 &  22.7\\
 5855.08 & 4.61 &  -1.48 & -30.21 & -4.58 & -4.59 &  24.4\\
 5916.25 & 2.45 &  -2.99 & -31.45 & -4.43 & -4.44 &  56.7\\
 6065.49 & 2.61 &  -1.53 & -31.41 & -4.51 & -4.52 & 132.1\\
 6082.71 & 2.22 &  -3.57 & -31.74 & -4.55 & -4.56 &  34.7\\
 6151.62 & 2.18 &  -3.30 & -31.58 & -4.52 & -4.53 &  51.1\\
 6200.32 & 2.61 &  -2.44 & -31.43 & -4.47 & -4.48 &  74.6\\ 
 6213.43 & 2.22 &  -2.48 & -31.58 & -4.58 & -4.59 &  86.0\\
 6229.23 & 2.85 &  -2.80 & -31.32 & -4.65 & -4.66 &  37.8\\
 6252.55 & 2.40 &  -1.69 & -31.52 & -4.52 & -4.53 & 135.4\\
 6393.61 & 2.43 &  -1.43 & -31.53 & -4.63 & -4.64 & 150.8\\
 6411.65 & 3.65 &  -0.60 & -30.38 & -4.57 & -4.57 & 154.7\\
 6421.35 & 2.28 &  -2.03 & -31.80 & -4.56 & -4.57 & 112.7\\
 6518.37 & 2.83 &  -2.46 & -31.37 & -4.59 & -4.60 &  64.0\\ 
Fe II &   &        &        &       &       &      \\
 4233.17 & 2.58 &  -1.97 & -32.01 & -4.64 & -4.64 & 132.1\\
 4508.29 & 2.84 &  -2.44 & -32.00 & -4.48 & -4.48 &  93.0\\ 
 4582.83 & 2.83 &  -3.18 & -32.03 & -4.54 & -4.54 &  57.7\\
 4620.52 & 2.82 &  -3.21 & -32.02 & -4.63 & -4.63 &  53.9\\
 4923.93 & 2.88 &  -1.26 & -32.03 & -4.69 & -4.69 & 196.1\\
 5018.44 & 2.88 &  -1.10 & -32.04 & -4.72 & -4.72 & 219.1\\
 5197.58 & 3.22 &  -2.22 & -32.02 & -4.58 & -4.58 &  85.4\\
 5264.81 & 3.22 &  -3.13 & -32.01 & -4.48 & -4.48 &  49.2\\
 5284.11 & 2.88 &  -3.11 & -32.04 & -4.57 & -4.57 &  60.3\\
 5325.55 & 3.21 &  -3.16 & -32.03 & -4.55 & -4.55 &  45.2\\
 5414.07 & 3.21 &  -3.58 & -32.02 & -4.52 & -4.52 &  29.8\\
 5425.26 & 3.20 &  -3.22 & -32.04 & -4.56 & -4.56 &  43.4\\
 5991.38 & 3.15 &  -3.55 & -32.05 & -4.58 & -4.58 &  32.1\\
 6239.95 & 3.89 &  -3.46 & -32.00 & -4.55 & -4.55 &  13.2\\
 6247.56 & 3.89 &  -2.30 & -32.00 & -4.58 & -4.58 &  55.0\\
 6369.46 & 2.89 &  -4.11 & -32.06 & -4.59 & -4.59 &  20.7\\
 6432.68 & 2.89 &  -3.57 & -32.07 & -4.56 & -4.56 &  43.2\\
 6456.38 & 3.90 &  -2.05 & -32.00 & -4.59 & -4.59 &  66.9\\
\enddata
\tablecomments{$gf$-values are from \citet{1979MNRAS.186..633B, 1982MNRAS.199...43B, 1982MNRAS.201..595B, 1991JOSAB...8.1185O}, and \citet{1991A&A...248..315B} for Fe~I and from \citet{mb2009} for Fe~II.}
\end{deluxetable}

\clearpage

\begin{deluxetable}{rllrrrrrrr}
\tabletypesize{\scriptsize}
\tablecaption{Final stellar parameters. \label{params}}
\tablehead{
\colhead{HD, } &  \colhead{ $\Teff$ } & \colhead{log g} &  \colhead{[Fe/H]} & \colhead{$\vt$}  & \colhead{mass} & \colhead{Fe I - Fe II }& \colhead{Fe I - Fe II } & \colhead{$N_{\rm Fe I}$}  & \colhead{$N_{\rm Fe II}$} \\
\colhead{ BD} &  \colhead{ K } & \colhead{} &  \colhead{} & \colhead{\kms}  & \colhead{$M_\odot$} & \colhead{LTE}& \colhead{NLTE} & \colhead{}  & \colhead{} 
}
\startdata
  \multicolumn{10}{c}{{\bf Benchmark stars}} \\
  19373~ &   6045 \scriptsize{$\pm$ 80} &  4.24 \scriptsize{$\pm$0.05} &  0.10 \scriptsize{$\pm$0.05}  & 1.2 & 1.11 &  0.06 \scriptsize{$\pm$0.09}  &  0.06 \scriptsize{$\pm$0.09}  &  26 &  15 \\
  22484~ &   6000 \scriptsize{$\pm$100} &  4.07 \scriptsize{$\pm$0.05} &  0.01 \scriptsize{$\pm$0.04}  & 1.1 & 1.18 & -0.02 \scriptsize{$\pm$0.07}  &  0.00 \scriptsize{$\pm$0.07}  &  27 &  16 \\
  22879~ &   5800 \scriptsize{$\pm$ 90} &  4.29 \scriptsize{$\pm$0.07} & -0.84 \scriptsize{$\pm$0.07}  & 1.0 & 0.75 & -0.06 \scriptsize{$\pm$0.08}  & -0.03 \scriptsize{$\pm$0.08}  &  23 &  14 \\
  30562~ &   5900 \scriptsize{$\pm$ 85} &  4.08 \scriptsize{$\pm$0.05} &  0.17 \scriptsize{$\pm$0.08}  & 1.3 & 1.12 &  0.02 \scriptsize{$\pm$0.09}  &  0.02 \scriptsize{$\pm$0.09}  &  26 &  16 \\
  34411~ &   5850 \scriptsize{$\pm$100} &  4.23 \scriptsize{$\pm$0.05} &  0.01 \scriptsize{$\pm$0.03}  & 1.2 & 1.10 & -0.01 \scriptsize{$\pm$0.05}  & -0.02 \scriptsize{$\pm$0.05}  &  26 &  15 \\
  49933~ &   6600 \scriptsize{$\pm$ 80} &  4.15 \scriptsize{$\pm$0.05} & -0.47 \scriptsize{$\pm$0.07}  & 1.7 & 1.11 & -0.05 \scriptsize{$\pm$0.08}  & -0.04 \scriptsize{$\pm$0.08}  &  26 &  15 \\
  59374~ &   5850 \scriptsize{$\pm$ 55} &  4.38 \scriptsize{$\pm$0.09} & -0.88 \scriptsize{$\pm$0.05}  & 1.2 & 0.75 & -0.06 \scriptsize{$\pm$0.09}  & -0.02 \scriptsize{$\pm$0.09}  &  24 &  15 \\
  59984~ &   5930 \scriptsize{$\pm$100} &  4.02 \scriptsize{$\pm$0.06} & -0.69 \scriptsize{$\pm$0.07}  & 1.4 & 0.89 & -0.06 \scriptsize{$\pm$0.10}  & -0.06 \scriptsize{$\pm$0.10}  &  23 &  18 \\
  64090~ &   5400 \scriptsize{$\pm$ 70} &  4.70 \scriptsize{$\pm$0.08} & -1.73 \scriptsize{$\pm$0.07}  & 0.7 & 0.62 & -0.02 \scriptsize{$\pm$0.09}  & -0.02 \scriptsize{$\pm$0.09}  &  22 &  11 \\
  69897~ &   6240 \scriptsize{$\pm$ 70} &  4.24 \scriptsize{$\pm$0.05} & -0.25 \scriptsize{$\pm$0.04}  & 1.4 & 1.15 & -0.06 \scriptsize{$\pm$0.07}  & -0.04 \scriptsize{$\pm$0.07}  &  28 &  15 \\
 84937~  &   6350 \scriptsize{$\pm$ 85}  &  4.09 \scriptsize{$\pm$0.08}  & -2.12 \scriptsize{$\pm$0.07} &  1.7 & 0.76 & -0.06 \scriptsize{$\pm$0.11}  &  0.00 \scriptsize{$\pm$0.12}  &  12  &   7 \\
  94028~ &   5970 \scriptsize{$\pm$130} &  4.33 \scriptsize{$\pm$0.08} & -1.47 \scriptsize{$\pm$0.04}  & 1.3 & 0.70 & -0.06 \scriptsize{$\pm$0.07}  & -0.04 \scriptsize{$\pm$0.07}  &  20 &  15 \\
 102870~ &   6170 \scriptsize{$\pm$ 80} &  4.14 \scriptsize{$\pm$0.04} &  0.11 \scriptsize{$\pm$0.06}  & 1.5 & 1.35 & -0.03 \scriptsize{$\pm$0.07}  & -0.01 \scriptsize{$\pm$0.07}  &  23 &  14 \\
 103095~ &   5130 \scriptsize{$\pm$ 65} &  4.66 \scriptsize{$\pm$0.08} & -1.26 \scriptsize{$\pm$0.08}  & 0.9 & 0.60 &  0.01 \scriptsize{$\pm$0.11}  &  0.01 \scriptsize{$\pm$0.11}  &  22 &   8 \\
 105755~ &   5800 \scriptsize{$\pm$ 55} &  4.05 \scriptsize{$\pm$0.09} & -0.73 \scriptsize{$\pm$0.05}  & 1.2 & 0.85 & -0.01 \scriptsize{$\pm$0.06}  &  0.00 \scriptsize{$\pm$0.06}  &  23 &  15 \\
 114710~ &   6090 \scriptsize{$\pm$ 80} &  4.47 \scriptsize{$\pm$0.05} &  0.06 \scriptsize{$\pm$0.06}  & 1.1 & 1.17 &  0.03 \scriptsize{$\pm$0.07}  &  0.04 \scriptsize{$\pm$0.06}  &  26 &  15 \\
 134169~ &   5890 \scriptsize{$\pm$ 80} &  4.02 \scriptsize{$\pm$0.07} & -0.78 \scriptsize{$\pm$0.07}  & 1.2 & 0.87 &  0.01 \scriptsize{$\pm$0.09}  &  0.06 \scriptsize{$\pm$0.09}  &  26 &  18 \\
 140283~ &   5780 \scriptsize{$\pm$ 55} &  3.70 \scriptsize{$\pm$0.07} & -2.46 \scriptsize{$\pm$0.07}  & 1.6 & 0.80 & -0.06 \scriptsize{$\pm$0.09}  &  -0.02 \scriptsize{$\pm$0.09}  &  20 &  14 \\
 142091~ &   4810 \scriptsize{$\pm$ 65} &  3.12 \scriptsize{$\pm$0.06} & -0.07 \scriptsize{$\pm$0.10}  & 1.2 & 1.16 &  0.05 \scriptsize{$\pm$0.13}  &  0.05 \scriptsize{$\pm$0.12}  &  20 &  13 \\
+66$^\circ$ 0268~ &   5300 \scriptsize{$\pm$ 80} &  4.72 \scriptsize{$\pm$0.11} & -2.06 \scriptsize{$\pm$0.15}  & 0.6 & 0.60 & -0.03 \scriptsize{$\pm$0.16}  & -0.03 \scriptsize{$\pm$0.16}  &  21 &   9 \\
  \multicolumn{10}{c}{{\bf Stars with spectroscopic parameters} } \\
  24289~ &   5980  &  3.71  & -1.94 \scriptsize{$\pm$0.17}  & 1.1 & 0.83 & -0.10 \scriptsize{$\pm$0.19}  & -0.03 \scriptsize{$\pm$0.19}  &  16 &  10 \\
  30743~ &   6450  &  4.20  & -0.44 \scriptsize{$\pm$0.07}  & 1.8 & 1.08 & -0.03 \scriptsize{$\pm$0.09}  &  0.01 \scriptsize{$\pm$0.09}  &  26 &  18 \\
  43318~ &   6250  &  3.92  & -0.19 \scriptsize{$\pm$0.08}  & 1.7 & 1.26 & -0.05 \scriptsize{$\pm$0.13}  & -0.01 \scriptsize{$\pm$0.09}  &  28 &  15 \\
  45067~ &   5960  &  3.94  & -0.16 \scriptsize{$\pm$0.06}  & 1.5 & 1.18 & -0.03 \scriptsize{$\pm$0.07}  & -0.01 \scriptsize{$\pm$0.07}  &  24 &  16 \\
  45205~ &   5790  &  4.08  & -0.87 \scriptsize{$\pm$0.03}  & 1.1 & 0.82 & -0.08 \scriptsize{$\pm$0.07}  & -0.05\scriptsize{$\pm$0.07} &  23 &  16 \\
  52711~ &   5900  &  4.33  & -0.21 \scriptsize{$\pm$0.05}  & 1.2 & 0.97 &  0.03 \scriptsize{$\pm$0.07}  &  0.05 \scriptsize{$\pm$0.07}  &  26 &  16 \\
  58855~ &   6410  &  4.32  & -0.29 \scriptsize{$\pm$0.05}  & 1.6 & 1.15 & -0.01 \scriptsize{$\pm$0.08}  &  0.01 \scriptsize{$\pm$0.07}  &  26 &  15 \\
  62301~ &   5840  &  4.09  & -0.70 \scriptsize{$\pm$0.04}  & 1.3 & 0.85 & -0.07 \scriptsize{$\pm$0.08}  & -0.04 \scriptsize{$\pm$0.08}  &  29 &  16 \\
  74000~ &   6225  &  4.13  & -1.97 \scriptsize{$\pm$0.07}  & 1.3 & 0.76 & -0.08 \scriptsize{$\pm$0.10}  & -0.02 \scriptsize{$\pm$0.10}  &  15 &   7 \\
  76932~ &   5870  &  4.10  & -0.98 \scriptsize{$\pm$0.05}  & 1.3 &      & -0.01 \scriptsize{$\pm$0.07}  &  0.03 \scriptsize{$\pm$0.08}  &  27 &  16 \\
  82943~ &   5970  &  4.37  &  0.19 \scriptsize{$\pm$0.04}  & 1.2 & 1.17 &  0.01 \scriptsize{$\pm$0.06}  &  0.01 \scriptsize{$\pm$0.06}  &  25 &  13 \\
  89744~ &   6280  &  3.97  &  0.13 \scriptsize{$\pm$0.03}  & 1.7 & 1.50 &  0.01 \scriptsize{$\pm$0.05}  &  0.02 \scriptsize{$\pm$0.06}  &  26 &  13 \\
  90839~ &   6195  &  4.38  & -0.18 \scriptsize{$\pm$0.05}  & 1.4 & 1.10 &  0.04 \scriptsize{$\pm$0.07}  &  0.06 \scriptsize{$\pm$0.07}  &  28 &  17 \\
  92855~ &   6020  &  4.36  & -0.12 \scriptsize{$\pm$0.03}  & 1.3 & 1.08 &  0.00 \scriptsize{$\pm$0.06}  &  0.02 \scriptsize{$\pm$0.05}  &  24 &  11 \\
  99984~ &   6190  &  3.72  & -0.38 \scriptsize{$\pm$0.04}  & 1.8 & 1.33 & -0.01 \scriptsize{$\pm$0.07}  &  0.03 \scriptsize{$\pm$0.07}  &  24 &  15 \\
 100563~ &   6460  &  4.32  &  0.06 \scriptsize{$\pm$0.08}  & 1.6 & 1.30 &  0.00 \scriptsize{$\pm$0.10}  &  0.02 \scriptsize{$\pm$0.10}  &  23 &  10 \\
 106516~ &   6300  &  4.44  & -0.73 \scriptsize{$\pm$0.06}  & 1.5 & 0.95 & -0.03 \scriptsize{$\pm$0.08}  &  0.00 \scriptsize{$\pm$0.08}  &  22 &  15 \\
 108177~ &   6100  &  4.22  & -1.67\scriptsize{$\pm$0.05}   & 1.1 & 0.72 & -0.08 \scriptsize{$\pm$0.13}  & -0.06 \scriptsize{$\pm$0.14}  & 16 &   5 \\
 110897~ &   5920  &  4.41  & -0.57 \scriptsize{$\pm$0.04}  & 1.2 & 0.85 &  0.03 \scriptsize{$\pm$0.05}  &  0.05 \scriptsize{$\pm$0.05}  &  29 &  18 \\
 115617~ &   5490  &  4.40  & -0.10 \scriptsize{$\pm$0.05}  & 1.1 & 0.93 & -0.04 \scriptsize{$\pm$0.08}  & -0.05 \scriptsize{$\pm$0.08}  &  26  &  13 \\
 134088~ &   5730  &  4.46  & -0.80 \scriptsize{$\pm$0.05}  & 1.1 & 0.75 & -0.02 \scriptsize{$\pm$0.07}  &  0.00 \scriptsize{$\pm$0.07}  &  25  &  15 \\
 138776~ &   5650  &  4.30  &  0.24 \scriptsize{$\pm$0.05}  & 1.3 & 0.93 &  0.02 \scriptsize{$\pm$0.10}  &  0.00 \scriptsize{$\pm$0.10}  &  21 &  11 \\
 142373~ &   5830  &  3.96  & -0.54 \scriptsize{$\pm$0.05}  & 1.4 & 0.95 & -0.02 \scriptsize{$\pm$0.07}  & -0.02 \scriptsize{$\pm$0.07}  &  28 &  17 \\
  --4$^\circ$ 3208~ &   6390 &  4.08  & -2.20 \scriptsize{$\pm$0.09}  & 1.4 & 0.77 & -0.08 \scriptsize{$\pm$0.11}  &  0.00 \scriptsize{$\pm$0.11}  &  16 &  11 \\
 --13$^\circ$ 3442~ &   6400 &  3.95  & -2.62 \scriptsize{$\pm$0.09}  & 1.4 & 0.79 & -0.14 \scriptsize{$\pm$0.11}  &  0.00 \scriptsize{$\pm$0.11}  &   8 &   5 \\
  +7$^\circ$ 4841~ &   6130  &  4.15  & -1.46 \scriptsize{$\pm$0.05}  & 1.3 & 0.79 &  0.00 \scriptsize{$\pm$0.09}  &  0.02 \scriptsize{$\pm$0.08}  &  21 &  17 \\
  +9$^\circ$ 0352~ &   6150  &  4.25  & -2.09 \scriptsize{$\pm$0.04}  & 1.3 &  0.70    & -0.01 \scriptsize{$\pm$0.07}  & 0.03 \scriptsize{$\pm$0.07}  &  15 &   6 \\
 +24$^\circ$ 1676~ &   6210  &  3.90  & -2.44 \scriptsize{$\pm$0.09}  & 1.5 & 0.78 & -0.06 \scriptsize{$\pm$0.12}  &  0.04 \scriptsize{$\pm$0.12}  &  12 &   5 \\
 +29$^\circ$ 2091~ &   5860  &  4.67  & -1.91 \scriptsize{$\pm$0.08}  & 0.8 & 0.70 & -0.05 \scriptsize{$\pm$0.10}  & -0.05 \scriptsize{$\pm$0.10}  &  18 &  10 \\
 +37$^\circ$ 1458~ &   5500  &  3.70  & -1.95 \scriptsize{$\pm$0.09}  & 1.0 &      & -0.07 \scriptsize{$\pm$0.11}  & -0.05 \scriptsize{$\pm$0.11}  &  21 &  14 \\
 G090-003 &  6010  &  3.90  & -2.04 \scriptsize{$\pm$0.06} & 1.3 & 0.78 & -0.06 \scriptsize{$\pm$0.10}  & -0.02 \scriptsize{$\pm$0.10}  &  17  &  12 \\
\enddata
\end{deluxetable}



\end{document}